\documentclass[%
 reprint,
 superscriptaddress,
 showpacs,preprintnumbers,
 amsmath,amssymb,
 aps,
 pra,
]{revtex4-1}

\DeclareMathAlphabet{\mathsfit}{T1}{\sfdefault}{\mddefault}{\sldefault}
\SetMathAlphabet{\mathsfit}{bold}{T1}{\sfdefault}{\bfdefault}{\sldefault}

\usepackage{graphicx}% Include figure files
\graphicspath{{fig/}}   %add figure path
\usepackage{dcolumn}% Align table columns on decimal point
\usepackage{bm}% bold math
\usepackage{makecell}
\usepackage{braket} %,mleftright}
\usepackage{siunitx}
\usepackage{color}
\usepackage{xr}
\usepackage{upgreek}
\usepackage[colorlinks]{hyperref}
\usepackage[capitalize]{cleveref}
\usepackage[caption=false]{subfig}
\usepackage{multirow} % for tables

\DeclareMathSymbol{\shortminus}{\mathbin}{AMSa}{"39}
\begin{document} 

\title{Quantum transport and localization in 1d and 2d tight-binding lattices}

\def\RLEaffil{Research Laboratory of Electronics, Massachusetts Institute of Technology, Cambridge, MA 02139, USA}
\def\LLaffil{MIT Lincoln Laboratory, Lexington, MA 02421, USA}
\def\Physaffil{Department of Physics, Massachusetts Institute of Technology, Cambridge, MA 02139, USA}
\def\EECSaffil{Department of Electrical Engineering and Computer Science, Massachusetts Institute of Technology, Cambridge, MA 02139, USA}
\def\Maryaffil{Laboratory for Physical Sciences, College Park, MD 20740, USA}
\def\Copenhagen{Center for Quantum Devices, Niels Bohr Institute, University of Copenhagen, 2100 Copenhagen, Denmark}

\author{Amir~H.~Karamlou}
\email{karamlou@mit.edu}
\affiliation{\RLEaffil}
\affiliation{\EECSaffil}
\author{Jochen~Braum\"uller}
\affiliation{\RLEaffil}
\author{Yariv~Yanay}
\affiliation{\Maryaffil}
\author{Agustin~Di~Paolo}
\affiliation{\RLEaffil}
\author{Patrick~Harrington}
\affiliation{\RLEaffil}
\author{Bharath~Kannan}
\affiliation{\RLEaffil}
\affiliation{\EECSaffil}
\author{David~Kim}
\affiliation{\LLaffil}
\author{Morten~Kjaergaard}
\affiliation{\RLEaffil}
\author{Alexander~Melville}
\affiliation{\LLaffil}
\author{Sarah~Muschinske}
\affiliation{\RLEaffil}
\affiliation{\EECSaffil}
\author{Bethany~Niedzielski}
\affiliation{\LLaffil}
\author{Antti~Veps\"al\"ainen}
\author{Roni~Winik}
\affiliation{\RLEaffil}
\author{Jonilyn~L.~Yoder}
\affiliation{\LLaffil}
\author{Mollie~Schwartz}
\affiliation{\LLaffil}
\author{Charles~Tahan}
\affiliation{\Maryaffil}
\author{Terry~P.~Orlando}
\affiliation{\RLEaffil}
\affiliation{\EECSaffil}
\author{Simon~Gustavsson}
\affiliation{\RLEaffil}
\author{William~D.~Oliver}
\email{william.oliver@mit.edu}
\affiliation{\RLEaffil}
\affiliation{\EECSaffil}
\affiliation{\LLaffil}
\affiliation{\Physaffil}

\date{\today}

\begin{abstract}

Particle transport and localization phenomena in condensed-matter systems can be modeled using a tight-binding lattice Hamiltonian. 
The ideal experimental emulation of such a model utilizes simultaneous, high-fidelity control and readout of each lattice site in a highly coherent quantum system. 
Here, we experimentally study quantum transport in one-dimensional and two-dimensional tight-binding lattices, emulated by a fully controllable $3 \times 3$ array of superconducting qubits. 
We probe the propagation of entanglement throughout the lattice and extract the degree of localization in the Anderson and Wannier-Stark regimes in the presence of site-tunable disorder strengths and gradients.
Our results are in quantitative agreement with numerical simulations and match theoretical predictions based on the tight-binding model. The demonstrated level of experimental control and accuracy in extracting the system observables of interest will enable the exploration of larger, interacting lattices where numerical simulations become intractable.

\end{abstract}

\maketitle

\section{Introduction}

\begin{figure*}[t]
\subfloat{\label{fig:concept}}
\subfloat{\label{fig:chip}}
\subfloat{\label{fig:pulse_sequence}}
\subfloat{\label{fig:1D_QRW}}
\subfloat{\label{fig:2D_QRW}}
\subfloat{\label{fig:qrw_distance}}
\includegraphics[width=\linewidth]{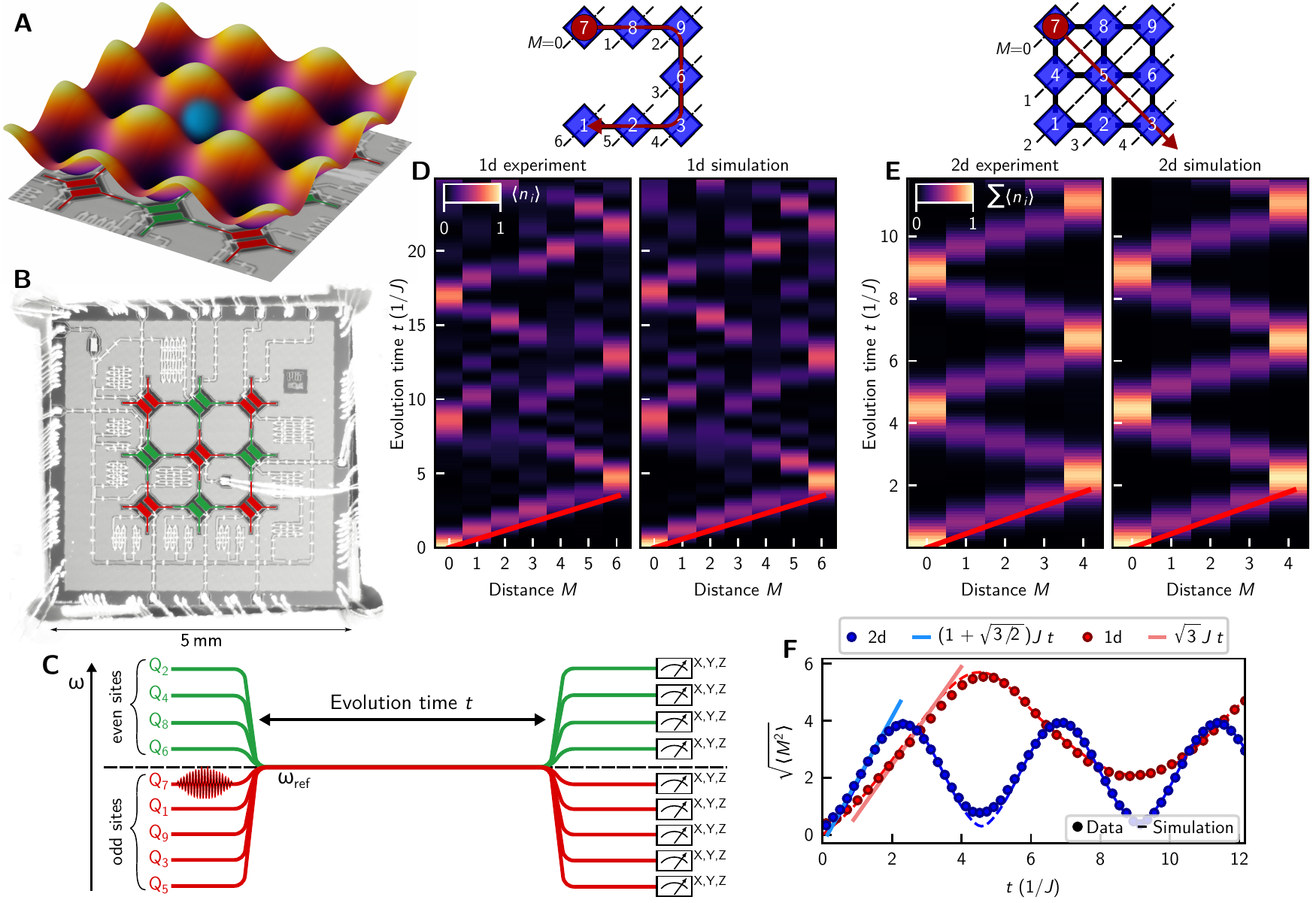}
\caption{
\textbf{Experimental concept} \textbf{(A)} 2d periodic potential emulated by the superconducting quantum circuit. \textbf{(B)} Optical image of the $3\times3$ superconducting qubit lattice used in the experiment. The capacitor pads of the transmons qubits are false colored. \textbf{(C)} Operation sequence of the experiments. Each qubit is initially detuned from $\omega_{\rm ref}$ and can be addressed individually, and simultaneously. They are brought on resonance to emulate $\hat{H}$ by evolving for time $t$. The system dynamics are frozen by once again detuning the qubits, and all nine qubits are measured simultaneously. \textbf{Quantum random walks} \textbf{(D)} A 1d quantum random walk with a single excitation (particle) initialized at the edge of the lattice (site 7). \textbf{(E)} A 2d quantum random walk with a single excitation (particle) initialized at the corner of the lattice (site 7). \textbf{(F)} The spatial propagation of the particle is measured by the root of the second moment of position ($\sqrt{\langle M^2 \rangle}$) in the 1d and 2d lattices. The excitation initialized at the edge of the 1d chain exhibits ballistic propagation in time and with propagation speed $v_g=\sqrt{3}J$ (red line), whereas an excitation initialized in the corner of a 2d lattice exhibits an increased propagation speed $v_g=(1+\sqrt{3/2})J$ (blue line).}
\label{fig:QRW}
\end{figure*}

Single-particle dynamics and transport in periodic solids is well described by the tight-binding model~\cite{Slater1954}.
This model is widely used to calculate the electronic band structure of condensed matter systems and to predict their transport properties, such as the conductance~\cite{Goringe1997, Cleri1993}. In the presence of a periodic lattice potential, the wavefunction of a given quantum particle overlaps neighboring lattice sites, leading to extended Bloch wavefunctions~\cite{Slater1954}. In the absence of lattice disorder, the particle propagation is ballistic and is described by a continuous-time quantum random walk~\cite{Kempe2003}. This is in contrast to classical diffusive transport, where the propagation is quadratically slower in time~\cite{Childs2001}.

The quantum nature of transport in lattices leads to the emergence of non-local correlations and entanglement between lattice sites. However, lattice inhomogeneity causes scattering and leads to quantum interference that tends to inhibit particle propagation, a signature of localization~\cite{Anderson1958,Emin1987}. The wavefunction of a localized particle rapidly decays away from its initial position, effectively confining the particle to a small region of the lattice.

Here, we study Anderson localization and Wannier-Stark localization in one-dimensional (1d) and two-dimensional (2d) tight-binding lattices. In the presence of random disorder of the lattice site energies, a particle wavefunction becomes spatially localized, known as Anderson localization~\cite{Anderson1958}. In the context of electrical transport, this phenomenon alters the transport properties, e.g., decreasing the conductance by increasing the degree of localization, ultimately leading to an insulating state. Alternatively, the particle can be localized in the presence of a potential gradient across the lattice, e.g., as created in the transport case by an external, static electric field, referred to as Wannier-Stark localization~\cite{Emin1987}. The potential gradient induces a position-dependent phase shift in the particle wavefunction and causes the particle to undergo periodic Bloch oscillations in a confined region~\cite{Hartmann2004}. While the dynamics of these oscillations are challenging to observe in naturally occurring solid-state materials~\cite{BenDahan1996}, they can be directly emulated and experimentally studied using engineered quantum systems.

Anderson localization has been experimentally realized in Bose-Einstein condensates~\cite{Billy2008,Roati2008}, atomic Fermi gases~\cite{Kondov2011}, and photonic lattices~\cite{Lahini2008,Schwartz2007}. Similarly, Bloch oscillations and Wannier-Stark localization have been observed in optical lattices~\cite{BenDahan1996,Preiss2015}, semiconductor superlattices~\cite{Feldmann1992}, and photonic lattices~\cite{Morandotti1999,Trompeter2006}. However, these demonstrations were limited to varying degrees by a lack of control and simultaneous readout of each individual site. Therefore, these experiments could not fully explore the different localization regimes and spatial dimensions of the tight-binding model in a single experimental instantiation.

%However, each of these demonstrations was limited to a specific localization regime due to the lack of full control over the potential of each individual site, and therefore cannot be utilized to explore the different localization regimes of the tight-binding model in different spatial dimensions in the single-particle limit.

%Superconducting quantum circuits are highly-controllable quantum systems that enable us to probe the properties of lattice models by emulating their Hamiltonian~\cite{Feynman1982,Georgescu2014}. The natural time evolution of the quantum circuits allows us to study the dynamics of quantum systems, while single- and two-qubit gate operations are used for state preparation and tomographic state readout~\cite{Roushan2017,Neill2018,Chiaro2019,Ma2019,Gong2021}.

Superconducting quantum circuits are highly-controllable quantum systems that enable us to probe the properties of lattice models. By engineering the Hamiltonian of the superconducting quantum circuit, we can emulate the dynamics of the lattice Hamiltonian, and we can use single- and two-qubit gate operations for state preparation and tomographic state readout~\cite{Roushan2017,Neill2018,Chiaro2019,Ma2019,Gong2021}. Site-selective qubit control, strong qubit-qubit interactions, and long coherence times relative to typical gate times combined with the capability of simultaneously applying gates and performing state readout on all sites make this a promising platform for studying models of information propagation, many-body entanglement, and quantum transport. 

In this work, we use a 9-qubit superconducting circuit to emulate the dynamics of single-particle quantum transport and localization in 1d and 2d tight-binding lattices. We probe the entanglement formed in the lattice as a result of the particle propagation. 
We then realize Anderson localization by implementing random disorder of the on-site energies with tunable strength, and extract the dependence of the propagation mean free path on the disorder strength in a regime with no analytical solution. 
%We implement tunable disorder strengths and extract the propagation mean free path in Anderson disorder regime that lack an analytical solution. 
Additionally, we study Wannier-Stark localization in the presence of isotropic and anisotropic potential gradients, and find close agreement between our experiments and theoretical predictions. Although performed on a small lattice that can still be simulated on a classical computer, our experiments demonstrate a platform for exploring larger, interacting systems where numerical simulations become intractable.

\section{Experimental setup}

Our device consists of a $3\times3$ array of capacitively coupled superconducting transmon qubits~\cite{Koch2007} (Fig.~\ref{fig:chip}), where qubit excitations correspond to particles in the lattice model. In the single-particle regime, the effective system Hamiltonian in the rotating frame with reference frequency $\omega_{\mathrm{ref}}$ is described by the tight-binding Hamiltonian:
\begin{equation}
    \hat{H}/\hbar= - \sum_{\langle i,j \rangle}J_{i,j} \hat{\sigma}_i^\dagger \hat{\sigma}_j + \sum_i \epsilon_i \hat{\sigma}_i^\dagger \hat{\sigma}_i
\label{eq:H}
\end{equation}
where $\hat{\sigma}_i^\dagger$ ($\hat{\sigma}_i$) are the two-level creation (annihilation) operators for each lattice site $i$. The first term in the Hamiltonian represents particle tunneling between neighboring sites with rate $J_{i,j}$, realized here with an average strength of $J/2\pi = \SI{8.1}{MHz}$ at $\omega_{\rm ref}=\SI{5.5}{GHz}$. The second term represents the particle occupation energy of sites $i$ with the corresponding site energies $\epsilon_i=\omega_i-\omega_{\rm ref}$, where $\omega_i$ is the fundamental transition frequency of the transmon at site $i$. By using flux-tunable transmons, the energy of each site can be individually set over the range $\SI{3.0}{GHz}\lesssim\omega_i/2\pi\lesssim\SI{5.5}{GHz}$. This tunability enables us to realize arbitrary energy landscapes and to isolate sub-lattices (e.g., one-dimensional chains) by detuning certain qubits from their neighbors so that they do not interact with the rest of the lattice. The transmons have an average anharmonicity of $U/2\pi= -\SI{244}{MHz}$, which corresponds to the on-site interaction energy of two particles occupying the same site. The system operates in the $J \ll |U|$ limit such that each lattice site can be occupied by at most a single particle, generally realizing the hard-core Bose-Hubbard model~\cite{Yanay2020,Braumuller2021}.

Our experiments feature simultaneous, site-resolved, single-shot dispersive qubit readout with an average qubit state assignment fidelity of $95\%$. In addition, we are able to control the energy of each site with an average precision $\langle \Delta \epsilon \rangle<(2\pi) \SI{200}{kHz}$ ($\approx J/40$). These features are crucial to accurately emulate the 1d and 2d tight-binding model with different lattice potential configurations.

\begin{figure*}[t]
\subfloat{\label{fig:concurrence_pairwise}}
\subfloat{\label{fig:distributed_concurrence}}
\subfloat{\label{fig:concurrence_average}}
\subfloat{\label{fig:concurrence_source_lattice}}
\subfloat{\label{fig:vn_entropy}}
\subfloat{\label{fig:MW_entanglement}}
\includegraphics[width=\linewidth]{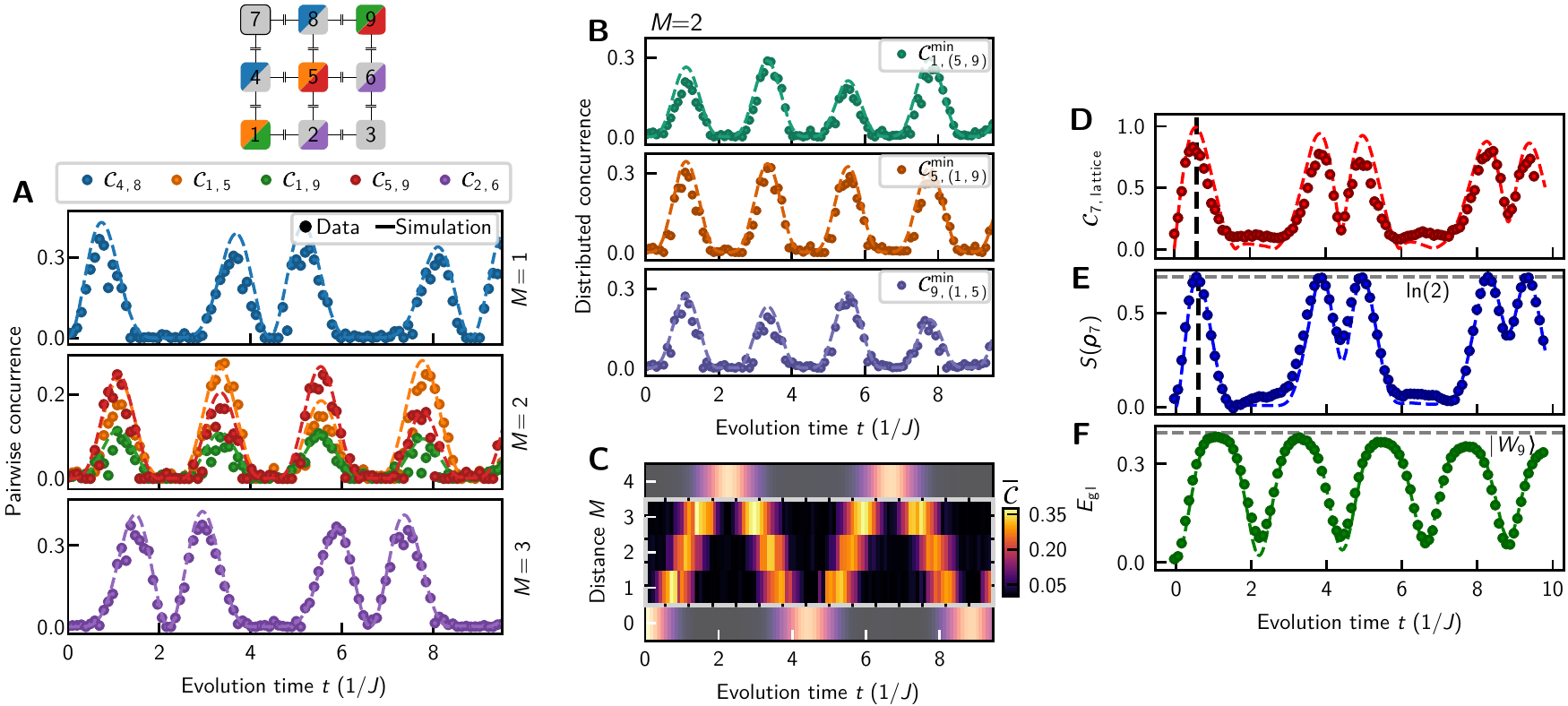}
\caption{
\textbf{Entanglement propagation in 2d lattice} \textbf{(A)} Evolution of the concurrence between pairs of sites at the same distance from the propagation source. \textbf{(B)} Distributed concurrence formed among the three qubits at distance $M=2$ from the particle initialization site. $\mathcal{C}^{\text{min}}_{i,(j,k)}$ represents the minimum concurrence between qubit $i$ and the sub-system consisting of qubits $j$ and $k$. \textbf{(C)} Average minimum concurrence formed between the sites along the diagonal symmetry axis of the lattice. We display the population on the sites at distance 0 and 4 as a reference (shaded regions). \textbf{(D)} Concurrence of the propagation source and the rest of the lattice. The source becomes maximally entangled with the remaining of the lattice at $t \simeq (2J)^{-1}$. \textbf{(E)} The von Neumann entropy of the propagation source (site 7), reaching a maximum value of $\rm{ln}(2)$ at time $t \simeq (2J)^{-1}$. \textbf{(F)} Global entanglement $E_{\rm gl}$ among the lattice sites. The  upper-bound of $E_{\rm gl}$ for a single-particle tight-binding model is set by the global entanglement of the 9-qubit $W$-state $\ket{W_9}$.}
\label{fig:entanglement}
\end{figure*}

\begin{figure*}[t]
\subfloat{\label{fig:population_localization_1d}}
\subfloat{\label{fig:distance_localization_1d}}
\subfloat{\label{fig:population_localization_2d}}
\subfloat{\label{fig:distance_localization_2d}}
\subfloat{\label{fig:E_gl_1d}}
\subfloat{\label{fig:E_gl_2d}}
\subfloat{\label{fig:PR_1D}}
\subfloat{\label{fig:PR_2D}}
\includegraphics{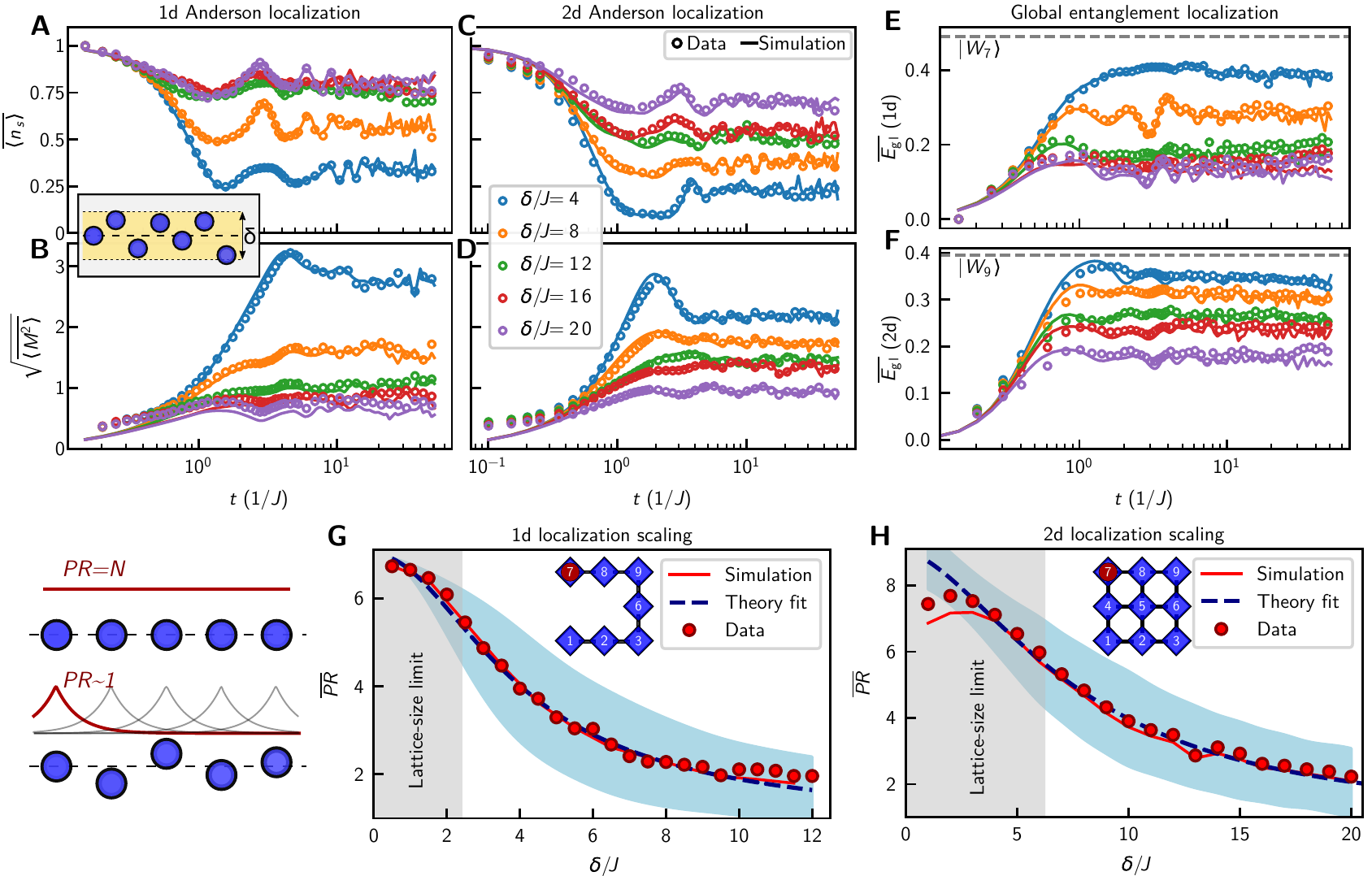}
\caption{
\textbf{Anderson Localization} Population at the propagation source \textbf{(A)} and particle root mean-squared distance from the source \textbf{(B)} for different 1d lattice disorders. Each point is averaged over 50 disorder realizations. Population at the propagation source \textbf{(C)} and particle mean-squared distance from the source \textbf{(D)} for different 2d lattice disorders. Each point is averaged over 50 disorder realizations. The global entanglement among the sites in the 1d lattice \textbf{(E)} and 2d lattice \textbf{(F)} measured at different disorder strengths averaged over $50$ realizations. \textbf{(G)} 1d Anderson localization participation ratio $\overline{\mathrm{PR}}$ scaling with disorder strength, obtained using $60$ random lattice realizations for each value of disorder strength $\delta$. The blue shaded region indicates the simulated statistical uncertainty interval ($\pm 1$ standard deviation) as a result of lattice randomization. \textbf{(H)} 2d Anderson localization $\overline{\mathrm{PR}}$ scaling with disorder strength, obtained using $180$ random lattice realization for each value of $\delta$.}
\label{fig:anderson_localization}
\end{figure*}

\section{Quantum Random Walk}

Quantum random walks (QRWs) are the quantum mechanical analog of classical random walks. The spatial propagation of the particle throughout the lattice, relative to its initial location, is quantified by the mean square displacement $\langle M^2 \rangle=\sum_i p_i M_i^2$, where $p_i$ is the probability of finding the particle on site $i$, and $M_i$ is the Manhattan (1-norm) distance between site $i$ and the initialization site. Quantum properties such as single-particle superposition and interference result in a
qualitative difference between classical and quantum random walks: a classical random walk propagates diffusively in time with $\sqrt{\langle M^2 \rangle} \propto \sqrt{t}$, whereas a QRW exhibits ballistic propagation with a mean-square displacement $\sqrt{\langle M^2 \rangle} \propto t$.

%We first study the particle propagation in a tight-binding system with uniform site energies ($\epsilon_i$=0). 
In order to experimentally observe particle propagation through QRWs, we first implement a tight-binding lattice with uniform site energies ($\epsilon_i$=0). 
We compare the respective propagation speeds for QRWs in a seven-site 1d  chain (Fig.~\ref{fig:1D_QRW}) and a 2d $3 \times 3$ lattice (Fig.~\ref{fig:2D_QRW}). We inject a particle at the end (corner) of the 1d (2d) lattice and observe its propagation by tracking the excitation numbers $\langle \hat{n}_i \rangle=\langle \hat{\sigma}_i^\dagger \hat{\sigma}_i \rangle$ on each lattice site versus evolution time $t$.

In the 1d case, the particle traverses the lattice and eventually reflects off the opposite end of the chain~\cite{Yan2019}. 
%
%The sites that are not located at either end of the chain are coupled to two neighbors, hence, the QRW can propagate in both directions. 
All intermediate sites are coupled to their two nearest neighbors, with the end sites being coupled to only one. 
Hence, the QRW propagates in both directions: an excitation is reemitted in both the forward and reverse directions (interim sites), or the direction is reversed (end sites).
Quantum interference between these multiple trajectories %as a result of the reflections at the lattice boundaries 
alters the particle wavepacket as it evolves in time. 
The resulting QRW pattern for a seven-site chain features a non-trivial evolution that agrees well with numerical simulation (Fig.~\ref{fig:1D_QRW}).
%Due to interference effects from these reflections at the lattice boundaries, the particle wave-packet shape is distorted, in agreement with numerical simulations (Fig.~\ref{fig:1D_QRW}).

In the 2d QRW, the particle propagates along its diagonal symmetry axis, analogous to a QRW on a binary tree~\cite{Childs2001}. Here, the particle position is defined by the sum of the site populations at a given Manhattan distance $M$ from the injection site. The quantum interference leading to the simple back-and-forth pattern as a result of reflection is exceptional and arises from the high-degree of symmetry present in a 3$\times$3 lattice (Fig.~\ref{fig:2D_QRW}). %, the 2d QRW is qualitatively similar to 
Such a high-degree of symmetry is similarly observed in 1d for a five-site chain~\cite{suppl}. 
%with modified effective coupling rates~\cite{suppl} .

To verify these experimental results, we perform Lindblad master equation simulations~\cite{suppl} based on the Hamiltonian in Eq.~\ref{eq:H}. We find excellent agreement between experimental data and numerical simulations by using the measured coupling strengths $J_{ij}$ between neighboring lattice sites $i,j$, and taking into account qubit relaxation and decoherence.

Both 1d and 2d QRWs exhibit ballistic propagation, and the propagation speed in 2d is faster than in 1d due to the constructive interference between multiple propagation paths leading to each site in 2d (Fig.~\ref{fig:qrw_distance}). The average group velocity $v_g$ of transport depends on the dimensionality of the lattice and the starting position of the particle with respect to the lattice boundaries. For a 1d QRW, a particle prepared at one end of the lattice initially propagates with $v_g=J$, because it interacts with only one other lattice site. At later times, prior to reflection from the other end, it reaches a steady state value of $v_g=\sqrt{3}J$, in agreement with theory for infinitely long chains~\cite{Hoyer2010}.
In contrast, we observe that a particle prepared in the corner of a $3\times 3$ lattice propagates with an initial group velocity $v_g=\sqrt{2}J$ (due to the coupling to its two nearest neighbors) and a steady state average velocity of $v_g=(1+\sqrt{3/2})J$~\cite{suppl} (see Fig.~\ref{fig:qrw_distance}).

During the QRW, certain sites become entangled as the particle propagates through the lattice, a phenomenon that underpins its intrinsic quantum character. We observe the emergence of entanglement during the system time evolution via non-local spatial correlations.
The amount of entanglement within a two-qubit sub-system, described by $\rho_{i,j}$, can be quantified using the pairwise concurrence~\cite{Wootters1998}:
\begin{equation}
    \mathcal{C}_{i,j} \equiv \mathcal{C}(\rho_{i,j})=\rm{max}\{0,\lambda_1-\lambda_2-\lambda_3-\lambda_4\}
\end{equation}
where $\{\lambda_i\}$ (in decreasing order) are the eigenvalues of the Hermitian matrix $R \equiv \sqrt{\sqrt{\rho_{i,j}} \tilde{\rho}_{i,j} \sqrt{\rho_{i,j}}}$. Here, $\tilde{\rho}=(\sigma_y \otimes \sigma_y) \rho^* (\sigma_y \otimes \sigma_y)$ is the spin-flipped density matrix obtained through complex conjugation and applying Pauli-Y operators ($\sigma_y$) to each qubit.
%where $\{\lambda_i\}$ (in decreasing order) are the eigenvalues of the Hermitian matrix $R \equiv \sqrt{\sqrt{\rho_{i,j}} (\sigma_y \otimes \sigma_y) \rho_{i,j}^* (\sigma_y \otimes \sigma_y) \sqrt{\rho_{i,j}}}$. Here, $\sigma_y$ is the Pauli-Y operator and $\rho_{i,j}^*$ is the complex conjugate of the density matrix. 
Concurrence is a monotone entanglement metric for two-qubit states that takes values between zero (product state) and one (maximally entangled). To reconstruct the two-qubit density matrix $\rho_{i,j}(t)$ and calculate the pairwise concurrence $\mathcal{C}_{i,j}(t)$, we perform two-qubit state tomography for various evolution times $t$ of the QRW. 

We measure the concurrence formed during a 2d QRW between site pairs at the same Manhattan distance from the particle initialization site (Fig.~\ref{fig:concurrence_pairwise}). Lattice sites at the same Manhattan distance become partially entangled as the particle traverses them, as indicated by an increase in the concurrence. This approach faithfully describes the quantum correlations in the sub-systems comprising of two sites ${M=1,3}$. However, the sub-system of sites at $M=2$ contains three qubits, and solely considering pairwise concurrence values does not fully capture the entanglement within this sub-system.
%We quantify the entanglement formed between the three sites at distance $M=2$ by extracting the concurrence within each pair of sites (middle panel in Fig.~\ref{fig:concurrence_pairwise}).

Using the pairwise concurrence values between the sites at $M=2$, we calculate the lower-bound of the distributed concurrence $\mathcal{C}_{i(j,k)}^{\rm min}$ for each site in this sub-system (Fig.~\ref{fig:distributed_concurrence}). For a state consisting of three qubits $(i,j,k)$, the concurrence $\mathcal{C}_{i(j,k)}$ between site $i$ and the system comprised of the two remaining sites $(j,k)$ can be lower-bounded using the pairwise concurrences~\cite{Coffman2000}:
\begin{equation}
    \mathcal{C}^2_{i(j,k)} \geq \mathcal{C}^2_{i,j}+\mathcal{C}^2_{i,k}.
    \label{concurrence_inequality}
\end{equation} 
Using the extracted pairwise and distributed concurrence values, we compute the average of the lower-bound concurrence $\Bar{\mathcal{C}}$ for sub-systems consisting of qubits that are at the same distance from the initial particle position (Fig.~\ref{fig:concurrence_average}). $\Bar{\mathcal{C}}$ is equal to the pairwise concurrence for sub-systems containing two qubits ($M=1,3$), and an average of $\mathcal{C}_{i(j,k)}^{\rm min}$ for the sub-system containing three qubits ($M=2$). The emergence of entanglement reflects the particle trajectory during the QRW and is in agreement with the measured quantum information propagation using out-of-time-ordered correlators~\cite{Braumuller2021, Mi2021}.

While the initially prepared single-particle state is separable from the rest of the lattice, the QRW, mediated by the nearest-neighbor interactions, creates and distributes entanglement to varying degrees throughout the lattice. 
%We quantify this entanglement using the measured pairwise concurrence values between the initialization site and each other site in the lattice~\cite{suppl}. 
For a system containing exactly a single particle, Eq.~\ref{concurrence_inequality} becomes an equality and can be generalized to a system with more than three sites~\cite{Coffman2000}. In our $3 \times 3$ lattice, the concurrence between the initialization site (site $7$) and the rest of the system $\mathcal{C}_{7,\rm{lattice}}$ can be extracted exactly using pairwise concurrence values~\cite{suppl}:
\begin{equation}
    \mathcal{C}^2_{7,\rm{lattice}} = \sum_{j^\prime \not = 7} \mathcal{C}^2_{7,j^\prime}.
\end{equation}
We observe that the quantum state of the initialization site becomes fully entangled with the larger system at time $t \simeq (2J)^{-1}$ (Fig.~\ref{fig:concurrence_source_lattice}). The coherent propagation leads to the fall and revival of the concurrence as the particle wavefunction evolves with time, with $\mathcal{C}_{7,\rm{lattice}}=0$ when site $7$ is in a pure ground or excited state. This behavior is in agreement with the evolution of the von Neumann entropy $S(\rho_7)= -\rm{tr}(\rho_7 \: \rm{ln} \: \rho_7)$ (Fig.~\ref{fig:vn_entropy}), where $\rho_7$ is the single-qubit density matrix of site $7$. The von Neumann entropy is a measure for the entanglement of the lattice site with its environment, and takes a value of $\rm{ln}(2)$ for a maximally entangled state. The dynamical revival of both $\mathcal{C}_{7,\rm{lattice}}$ and $S(\rho_7)$ indicates that site $7$ becomes predominantly %coherently 
entangled with the rest of the lattice, rather than with the uncontrolled environmental degrees of freedom %in the environment %due 
related to decoherence.

Without the need for full quantum state tomography, we probe the global entanglement $ E_{\rm gl}$ in the many-body system, beyond the non-local correlations formed between sub-systems, by extracting the average purity of all $N$ sites~\cite{Meyer2002, Amico2008}
\begin{equation}
    E_{\rm gl}=2-\frac{2}{N}\sum_{j=1}^N \rm{tr}(\rho^2_j),
\end{equation}
where $\rho_j$ is the reduced density matrix describing site $j$. In Fig.~\ref{fig:MW_entanglement}, we observe that the global entanglement reaches a maximum value when the particle is not concentrated on either corner of the lattice, but rather extends across all lattice sites. In our 2d lattice, $E_{\rm gl}$ is upper-bounded by the global entanglement in the 9-qubit $W$-state $\ket{W_9}$~\cite{Brennen2003}, where $\ket{W_N}=\frac{1}{\sqrt{N}} \sum_\pi \ket{\pi (0...01)}$ is the superposition of all single-particle state permutations.

\section{Anderson Localization}

We have so far explored quantum transport in lattices with uniform site energies. However, particle propagation is significantly impacted by the introduction of random disorder in the site energies, leading to Anderson localization~\cite{Anderson1958}. In the presence of disorder in 1d and 2d lattices, the particle wavefunction becomes spatially localized, regardless of the disorder strength in the thermodynamic limit~\cite{Lee1985}. In our experiments, we emulate the Anderson localization regime by introducing disorder with strength $\delta$, where site energies are randomly sampled from a  uniform distribution $\epsilon_i \in [-\delta/2,\delta/2]$. As a result, wavefunction scattering causes the particle to lose phase coherence on the length scale of the mean-free path $l$. While an analytical form for the mean free path can be derived in the limit of weak ($\delta \ll J$) and strong ($\delta \gg J$) disorder~\cite{Casati1992,Chuang2016}, there are no known analytical expressions in the intermediate disorder regime~\cite{Varga1994}.

We examine the degree of localization by measuring the increase in average population $\overline{\langle n_s \rangle}$ of the initially prepared lattice site and the decrease in particle spread $\sqrt{\overline{\langle M^2 \rangle}}$ for increasing disorder strengths $\delta$, each averaged over $50$ random lattice realizations.
As the disorder strength increases, a higher steady-state population remains on the source site as a result of the tight-binding interaction (Fig.~\ref{fig:population_localization_1d} and~\ref{fig:population_localization_2d}), while the average particle spread away from the source decreases (Fig.~\ref{fig:distance_localization_1d} and~\ref{fig:distance_localization_2d}) for both 1d and 2d lattices. For a given disorder strength, the particle is more confined in a 1d lattice compared to a 2d lattice; in 2d, the probability of transport is relatively greater in the presence of disorder as propagation occurs along multiple paths between two sites. As a result of localization caused by disorder, the emergence of multipartite entanglement in the tight-binding lattice is inhibited: in Fig.~\ref{fig:E_gl_1d} and Fig.~\ref{fig:E_gl_2d}, we report a decrease in the steady state value for the average global entanglement of the system, $\overline{E_{\rm gl}}$, with increasing disorder strength, due to spatial localization of the particle wavefunction. With one excitation, the confinement effects for seven-site chain causes a relatively greater deviation from the maximum value ($E_{\rm gl}$ of $\ket{W_7}$) compared to the 2d lattice.
%As a result of stronger propagation confinement in a 1d lattice, we observe a greater deviation from the maximum $E_{\rm gl}$ for a seven-site system with a single particle ($\ket{W_7}$).

We quantify the localization of the wavefunction in a lattice with $N$ sites using the participation ratio $\mathrm{PR}$, defined as~\cite{VanNieuwenburg2019} 
\begin{equation}
    \mathrm{PR}(\psi)=\left(\sum^N_i| \psi_i|^4\right)^{-1}=\left(\sum^N_i| n_i|^2\right)^{-1}.
\end{equation}
If the particle wavefunction is completely delocalized, then $\mathrm{PR}=N$, whereas $\mathrm{PR}=1$ for a wavefunction fully localized to a single site. In our experiments, $\mathrm{PR}$ for each lattice realization is calculated using the time-averaged population on each site after the tight-binding model dynamics reach steady state~\cite{VanNieuwenburg2019}. In order to reduce the impact of qubit relaxation on these measurements, we use the averaged site populations between $\SI{100}{ns}$ and $\SI{400}{ns}$ ($5/J\lesssim t \lesssim 20/J$) to calculate $\mathrm{PR}$.
We experimentally extract the participation ratio $\overline{\mathrm{PR}}$ averaged over different random lattice realizations, at different disorder strengths for both a 1d and 2d lattice. As $\delta$ increases, the particle wavefunction becomes more spatially confined and $\overline{\mathrm{PR}}$ decreases. For weak disorder strengths, we find the localization length to be larger than the lattice size (gray region in Fig.~\ref{fig:PR_1D} and Fig.~\ref{fig:PR_2D}), imposing a limitation on our experiments caused by boundary effects. 

In a finite 1d lattice, 
%the value of PR 
the participation ratio is related to the localization length $\xi_{\rm 1d}$ through the expression $\mathrm{PR}(\xi_{\rm 1d}) =  \rm{coth(1/\xi_{\rm 1d})} \: \rm{tanh} (N/\xi_{\rm 1d})$~\cite{suppl}. The 1d localization length $\xi_{\rm 1d}$ scales directly with the mean free path $l$ as $\xi_{\rm 1d} = l$~\cite{Lee1985}. We experimentally extract the dependence of the mean free path on the disorder strength, taking the form $l=a(J/\delta)^{\gamma}$, in our seven-site chain by realizing $60$ random lattice disorders for different disorder strengths and calculating $\overline{\mathrm{PR}}$ (Fig.~\ref{fig:PR_1D}). In the 1d case, we find $\gamma=1.0 \pm 0.03$ and $a=17.0 \pm 1.05$ (in units of the lattice spacing) through fitting.

In a 2d $n\times n$ lattice, 
%
%$\overline{\mathrm{PR}}$ is related to the localization length through 
the participation ratio is related to the localization length through the expression
$\mathrm{PR}(\xi_{\rm 2d}) =  \rm{coth}^2(1/\xi_{\rm 2d}) \: \rm{tanh}^2 (n/\xi_{\rm 2d})$~\cite{suppl}, where the localization length takes the form $\xi_{\rm 2d}= l\: e^{\frac{\pi}{2} k\:l}$ with a lattice-dependent factor $k$~\cite{Lee1985}. For weak disorder, the 2d $\overline{\mathrm{PR}}$ scales exponentially with the mean free path. Consequently, the scaling for small values of $\delta$ is difficult to observe in our small lattice, and hence the value of $k$ cannot be obtained with high accuracy. We extract the 2d $\overline{\mathrm{PR}}$ in our $3 \times 3$ lattice (Fig.~\ref{fig:PR_2D}) over $180$ random lattice realizations for different disorder strengths, and for large values of $\delta$ we find the parameters for the mean free path $\gamma=0.80 \pm 0.02$ and $a=12.98 \pm 0.60$ (in units of the lattice spacing). In a 2d lattice, we notice that the extracted mean free path exhibits a weaker dependence on the disorder strength, %which is 
consistent with our earlier observations.

\section{Wannier-Stark Localization}

\begin{figure*}[t]
\subfloat{\label{fig:Bloch_1d_profile}}
\subfloat{\label{fig:T_B_1d}}
\subfloat{\label{fig:max_distance_1d}}
\subfloat{\label{fig:Bloch_2d_profile}}
\subfloat{\label{fig:T_B_2d}}
\subfloat{\label{fig:max_distance_2d}}
\subfloat{\label{fig:Bloch_asym_profile}}
\subfloat{\label{fig:T_B_asym}}
\subfloat{\label{fig:max_distance_asym}}
\includegraphics{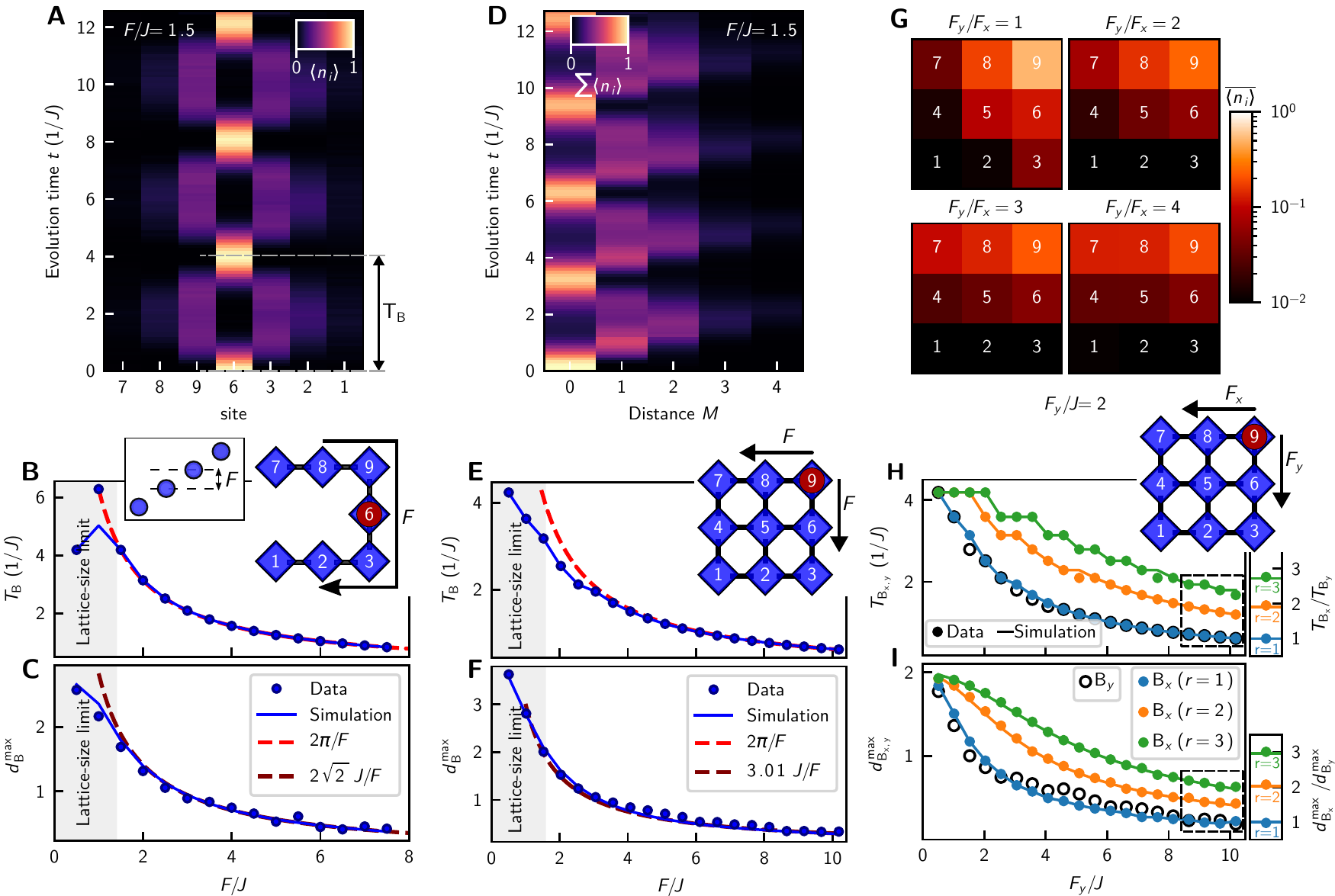}
\caption{
\textbf{Wannier-Stark Localization} \textbf{(A)} Exemplary 1d Bloch oscillation with a particle initialized at the center of the lattice. The particle undergoes a spatially periodic breathing motion before re-converging at the initialization site. \textbf{(B)},\textbf{(C)} 1d Bloch oscillation period $T_{\rm B}$ and maximum particle spread $d_{\rm B}^{\rm max}$ scaling with the potential gradient strength $F$. \textbf{(D)} Exemplary 2d Bloch oscillation with an isotropic potential gradient with a particle initialized at the corner of the lattice. The particle undergoes periodic Bloch oscillations along the diagonal symmetry axis of the lattice. \textbf{(E)},\textbf{(F)} Isotropic 2d Bloch oscillation period $T_{\rm B}$ and maximum particle spread $d_{\rm B}^{\rm max}$ depending on $F$. \textbf{(G)} Time-averaged population $\overline{\langle n_i \rangle}$ on each site for different anisotropy ration $r=F_y/F_x$. As $r$ increases, the propagation gets skewed towards the horizontal axis. \textbf{(H)},\textbf{(I)} $T_{\rm B}$ and $d_{\rm B}^{\rm max}$ along the horizontal ($B_x$) and vertical ($B_y$) direction of the lattice of anisotropic fields at different values of $r$. Inset on the right side of the figure show the values of $T_{\rm{B}_x}/T_{\rm{B}_y}$ and $d^{\rm{max}}_{\rm{B}_x}/d^{\rm{max}}_{\rm{B}_y}$ obtained from our experiments and numerical simulations highlighted using a dashed box.}
\label{fig:stark_localization}
\end{figure*}

A particle in the tight-binding lattice is localized also in the presence of a static electric field, which creates a potential gradient across lattice site energies~\cite{Emin1987}. We emulate  the effect of an electric field by creating a gradient in the lattice site energies $\epsilon_i$. The resulting potential gradient causes the particle to become spatially confined due to the emergence of a band-gap in the lattice band structure~\cite{BenDahan1996}. This phenomenon is referred to as Wannier-Stark localization~\cite{Emin1987,Preiss2015,Guo2021}. The effective field creates a linear gradient $\vec{F}$ along the lattice axis, where $\vec{F}=\nabla  \epsilon_i$, and causes the particle to undergo periodic Bloch oscillations.

We demonstrate Bloch oscillations in a 1d chain for the field strength $F/J=1.5$ ($F=|\vec{F}|$) observing a spatially periodic breathing motion with a probability up to $99\%$ of the particle reviving at the initialization site (Fig.~\ref{fig:Bloch_1d_profile}). By repeating the experiment for different values of $F$, we observe that a particle initialized in the center of the lattice oscillates with a Bloch period $T_{\rm B}=2\pi /F$ (Fig.~\ref{fig:T_B_1d}) and a maximum particle spread of $d^{\rm{max}}_{\rm B}=\mathrm{max} (\sqrt{\langle M^2 \rangle})=2\sqrt{2}J/F$ (Fig.~\ref{fig:max_distance_1d}) from the initialization site, in agreement with theory~\cite{Hartmann2004}. The finite lattice size causes the boundaries to have a notable effect on the oscillation for small values of $F$ (gray region of Fig.~\ref{fig:T_B_1d} and Fig.~\ref{fig:max_distance_1d}), namely the decrease in the oscillation period and limiting the particle spread. We observe a similar periodic motion along the diagonal symmetry axis of the lattice in 2d (Fig.~\ref{fig:Bloch_2d_profile}) for an isotropic field ($F=F_x=F_y$). The Bloch period of a particle initialized in the corner of a 2d lattice with this symmetric gradient exhibits the same scaling with $F$ as in 1d. We find that the maximum particle spread in a 2d lattice scales as $d^{\rm{max}}_{\rm B}=\mathrm{max} (\sqrt{\langle M^2 \rangle})=3.01 \pm 0.01 \: J/F$, where we obtained the value 3.01 from a data fitting procedure. Wannier-Stark localization is less pronounced in 2d compared to 1d for the same potential gradient, similar to the trend we observe in the Anderson localization regime and in agreement with theoretical predictions~\cite{Witthaut2004}.  

We also explore the effect of non-isotropic fields in a 2d lattice by independently controlling the potential gradients along the horizontal ($F_x$) and the vertical ($F_y$) axes, respectively. In Fig.~\ref{fig:Bloch_asym_profile}, we show the time-averaged population of each site during the first $\SI{500}{ns}$ ($\approx 25/J$) of evolution as a result of different field ratios $r=F_y/F_x$. In the presence of the same field along both lattice axes ($r=1$), the particle is localized along the diagonal symmetry axis of the lattice (Fig.~\ref{fig:Bloch_asym_profile} top left). As $r$ increases by decreasing $F_x$ while keeping $F_y$ fixed, the particle becomes less localized along the $F_x$-direction, %and thereby 
skewing the propagation direction. 

In the absence of boundary effects, the relation between Bloch oscillation periods $T_{\rm{B}_x}$ and $T_{\rm{B}_y}$ along each lattice axis depends on the net field direction, namely $T_{\rm{B}_x}=r \: T_{\rm{B}_y}$ (Fig.~\ref{fig:T_B_asym}). 
For rational values of $r$, the total 2d Bloch period $T_{\rm{B}}$ is the least common multiple of the one-dimensional Bloch periods along each axis~\cite{Witthaut2004}. 
A similar relationship holds for the maximum particle spread in each direction $d^{\rm{max}}_{\rm{B}_x}=r \: d^{\rm{max}}_{\rm{B}_y}$ (Fig.~\ref{fig:max_distance_asym}). %
In our experiments, we investigate these relations by measuring $T_{B_x}$ (see Fig.~\ref{fig:T_B_asym}) and $d^{\textrm{max}}_{B_x}$ (see Fig.~\ref{fig:max_distance_asym}) for $r=1,2,3$ (solid circles) while varying $F_y$. 
The data match simulation (solid lines) of the $3\times3$ system very well. 
%and extracting the oscillation period and maximum spread along the horizontal ($\rm{B}_x$) and vertical ($\rm{B}_y$) axes. 
%
We additionally measured $T_{B_y}$ and $d^{\textrm{max}}_{B_y}$, which are nominally unchanged for $r=1,2,3$, and present the values measured for isotropic potential gradients (open circles). 
This enables us to obtain experimental estimates for $r$ by taking the ratio between the experimentally measured Bloch periods $T_{\rm{B}_x}/T_{\rm{B}_y}$ (Fig.~\ref{fig:T_B_asym}, right panel) and the maximum spread $d^{\rm{max}}_{\rm{B}_x}/d^{\rm{max}}_{\rm{B}_y}$ (Fig.~\ref{fig:max_distance_asym}, right panel). The ratios for each $r$ are estimated from highlighted points in Figs.~\ref{fig:T_B_asym} and~\ref{fig:max_distance_asym}. 
%For the Bloch-period ratios, the slight deviation between the experimentally inferred values and the actual values of $r$ (particularly for $r=2,3$) can be attributed to the boundary effects of our lattice, and matches the numerical simulations for our $3\times3$  system. 
%For the particle-spread ratios, the deviation is a bit larger and related primarily to the small amount of noise in the $d^{\textrm{max}}_{B_y}$ measurement (open circles, Fig.~\ref{fig:max_distance_asym}).
%Due to the limited lattice size, we observe a slight deviation in our experimental data and numerical simulations from these predictions as the value of $r$ becomes larger.

\section{Conclusion}

In this work, we have emulated quantum particle propagation and localization in 1d and 2d tight-binding lattices using a $3\times3$ array of superconducting qubits. We have measured the coherent dynamics of different entanglement metrics, such as concurrence and the von Neumann entropy, during quantum transport using simultaneous control and readout. We have further investigated localization in different regimes of the tight-binding model, with random disorder resulting in Anderson localization and with a potential gradient causing Wannier-Stark localization and Bloch oscillations. We have studied Anderson localization in a disorder regime that lacks an analytical form, and used our data to extract the dependence of the propagation mean free path in finite 1d and 2d lattices on the disorder strength. We have measured the degree of Wannier-Stark localization as a result of different potential gradients for both isotropic and anisotropic potentials, and have observed quantitative agreement in the properties of Bloch oscillations with theoretical predictions. The demonstrated degree of control in realizing different regimes of the tight-binding model in 1d and 2d serves as a blueprint for exploring larger and strongly interacting quantum lattices in the many-body localization regime~\cite{VanNieuwenburg2019, Sierant2020, Khemani2020} or that contain a time-dependent Hamiltonian leading to interesting phenomena such as quantum scars~\cite{Mukherjee2020} topological order~\cite{Goldman2014} and the breakdown of the Magnus expansion~\cite{DAlessio2014}.
\\
\vspace{0.2in}
\section*{Acknowledgments}
The authors are grateful to F.~Vasconcelos, and S.~Lloyd for insightful discussions.
AHK acknowledges support from the NSF Graduate Research Fellowship Program. 
This research was funded in part by the U.S.~Army Research Office Grant W911NF-18-1-0411 and the Assistant Secretary of Defense for Research \& Engineering under Air Force Contract No. FA8721-05-C-0002. Opinions, interpretations, conclusions, and recommendations are those of the authors and are not necessarily endorsed by the United States Government.

\bibliographystyle{prlike}

\bibliography{refs}

\end{document}

\title{Supplementary material for ``Quantum transport and localization in 1d and 2d tight-binding lattices''}

\def\RLEaffil{Research Laboratory of Electronics, Massachusetts Institute of Technology, Cambridge, MA 02139, USA}
\def\LLaffil{MIT Lincoln Laboratory, Lexington, MA 02421, USA}
\def\Physaffil{Department of Physics, Massachusetts Institute of Technology, Cambridge, MA 02139, USA}
\def\EECSaffil{Department of Electrical Engineering and Computer Science, Massachusetts Institute of Technology, Cambridge, MA 02139, USA}
\def\Maryaffil{Laboratory for Physical Sciences, College Park, MD 20740, USA}

\author{Amir~H.~Karamlou}
\email{karamlou@mit.edu}
\affiliation{\RLEaffil}
\affiliation{\EECSaffil}
\author{Jochen~Braum\"uller}
\affiliation{\RLEaffil}
\author{Yariv~Yanay}
\affiliation{\Maryaffil}
\author{Agustin~Di~Paolo}
\affiliation{\RLEaffil}
\author{Patrick~Harrington}
\affiliation{\RLEaffil}
\author{Bharath~Kannan}
\affiliation{\RLEaffil}
\affiliation{\EECSaffil}
\author{David~Kim}
\affiliation{\LLaffil}
\author{Morten~Kjaergaard}
\affiliation{\RLEaffil}
\author{Alexander~Melville}
\affiliation{\LLaffil}
\author{Sarah~Muschinske}
\affiliation{\RLEaffil}
\affiliation{\EECSaffil}
\author{Bethany~Niedzielski}
\affiliation{\LLaffil}
\author{Antti~Veps\"al\"ainen}
\author{Roni~Winik}
\affiliation{\RLEaffil}
\author{Jonilyn~L.~Yoder}
\affiliation{\LLaffil}
\author{Mollie~Schwartz}
\affiliation{\LLaffil}
\author{Charles~Tahan}
\affiliation{\Maryaffil}
\author{Terry~P.~Orlando}
\affiliation{\RLEaffil}
\affiliation{\EECSaffil}
\author{Simon~Gustavsson}
\affiliation{\RLEaffil}
\author{William~D.~Oliver}
\email{william.oliver@mit.edu}
\affiliation{\RLEaffil}
\affiliation{\EECSaffil}
\affiliation{\LLaffil}
\affiliation{\Physaffil}

\date{\today}

\maketitle

\tableofcontents

\section{Sample and experimental setup}

\begin{figure*}
\includegraphics{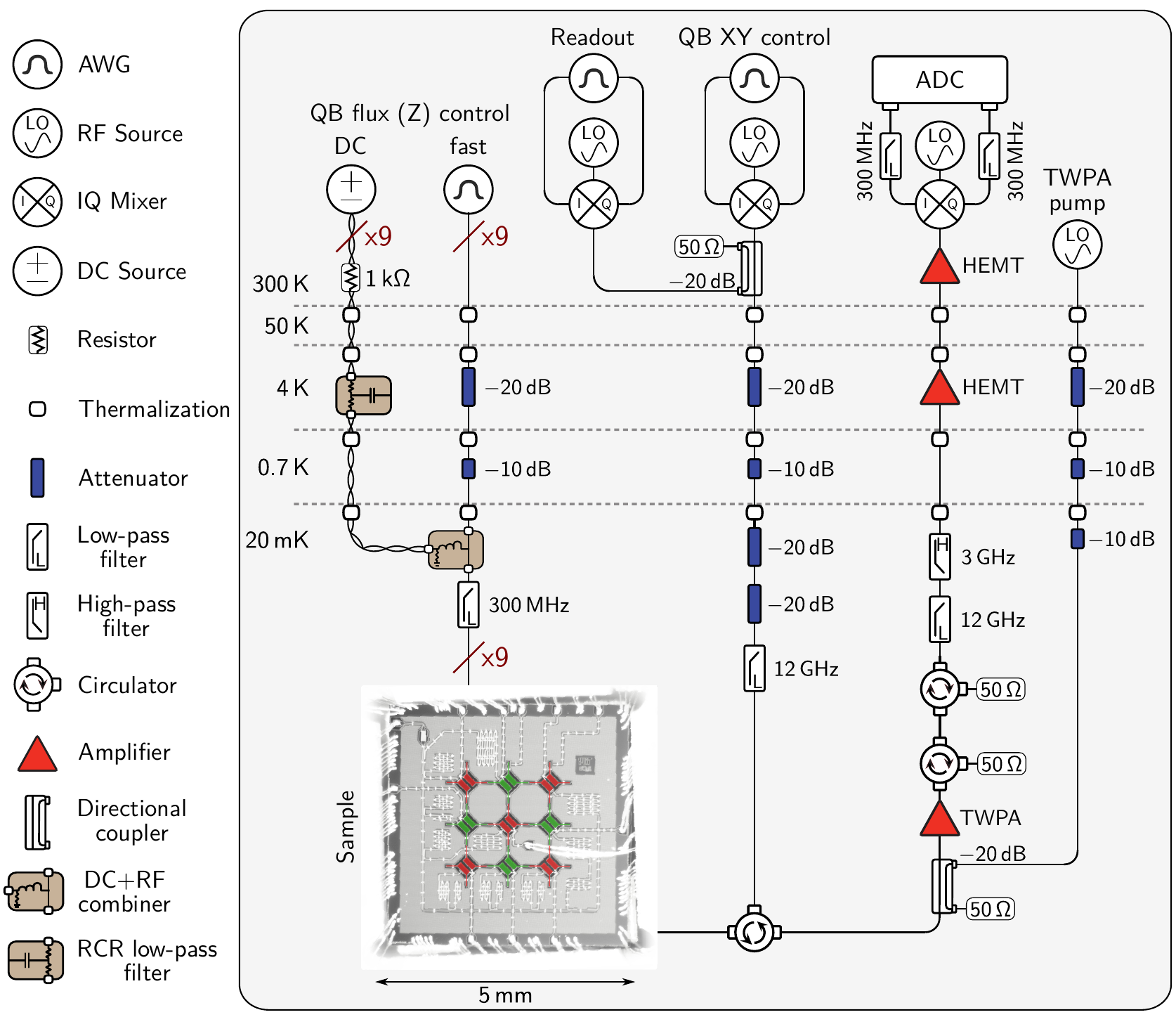}
\caption{
\textbf{Sample and experimental setup}
}
\label{fig:setup}
\end{figure*}

The sample used in our experiments (inset in Fig.~\ref{fig:setup}) is a two-dimensional (2d) lattice of capacitively coupled transmon qubits~\cite{Koch2007} with nearest-neighbor qubit-qubit interactions of average strength $J/2\pi=(8.1\pm 0.2)\,\mathrm{MHz}$. The floating capacitor pads of the transmons are arranged such that the coupling phases add up to zero within a closed loop of qubits, yielding vanishing effective gauge fields. The next-nearest-neighbor interaction is strongly suppressed and is smaller than $\sim J/75$~\cite{Braumuller2021} with an anisotropy caused by the capacitance configuration of the qubit lattice.

Each qubit is coupled to an individual coplanar resonator for dispersive qubit state readout~\cite{Blais2004,Krantz2019}. We use $\lambda/4$ readout resonators for all qubits along the edge of the lattice to minimize their footprint on the sample, and an `open-terminated' $\lambda/2$ readout resonator for the central qubit, which crosses a corner qubit at its voltage node in order to minimize the parasitic coupling~\cite{Braumuller2021}. The readout resonators couple to a single Purcell filter resonator with a bandwidth of about $\SI{0.54}{GHz}$ and a resonance frequency of $\SI{7.3}{GHz}$, facilitating simultaneous readout of the states of all nine qubits at an average qubit state assignment fidelity of $95\%$. The effective readout resonator linewidths are in the range of $(1-3)\,\mathrm{MHz}$ and our readout integration time is $\SI{800}{ns}$. The relatively low quality factor of the Purcell filter enables us to drive the qubits through the common readout line. Each qubit is coupled to a flux bias line used for both static (DC) and pulsed (`fast') flux control with a bandwidth of $\sim\SI{400}{MHz}$, set by by the arbitrary waveform generator. The average measured mutual inductance between each SQUID loops and its respective flux bias line is $\SI{0.5}{pH}$, matching our design value. We find dc flux cross-talk up to $30\%$ due to the long bond wire reaching the central qubit for flux control, which we compensate via flux cross-talk calibration. We perform separate dc and fast flux cross-talk calibrations, and utilize a learning-based approach for optimizing the cross-talk matrices, followed by qubit spectroscopy to further fine-tune the qubit frequencies~\cite{Braumuller2021}, yielding qubit frequency accuracies of $\SI{0.2}{MHz}\approx J/40$. In order to cancel distortions in the fast flux control pulses, we perform transient calibrations for each flux lines individually. For more information on the setup calibration see~\cite{Braumuller2021}.

\begin{table}
\centering
\caption{\textbf{Sample parameters}. We show the maximum transmon transition frequencies $\omega_{\mathrm{q}}^{\rm max}$ at the upper flux insensitive point, the qubit anharmonicities $U$ (measured at $\omega_{\mathrm{q}}^{\rm max}$), the readout resonator frequencies $\omega_{\mathrm{r}}$, the probabilities $f_{ij}$ of measuring the qubit in state $i$ after preparing it in state $j$, the readout assignment fidelity $(f_{\mathrm{gg}}+f_{\mathrm{ee}})/2$, and the average measured $T_1^{\rm avg}$ at qubit frequencies around the bias point used in experiment.}
\vspace{8pt}
\label{tab:sample_parameters}
{\renewcommand{\arraystretch}{1.7}   %modifying vertical spacing in tabular
\begin{tabular}{ p{4.5cm}  p{1cm} p{1cm}  p{1cm}  p{1cm} p{1cm} p{1cm}  p{1cm}  p{1cm} p{1cm}   }
\toprule
   &    Q1  &          Q2  &          Q3  &          Q4  &          Q5  &          Q6  &          Q7  &          Q8  &          Q9  \\
\hline
$\omega_{\mathrm{q}}^{\rm max}/2\pi$ (GHz) &  5.712 & 5.771 & 5.788 & 5.707 & 5.822 & 5.626 & 5.528 & 5.673 & 5.722  \\

$U/2\pi$ (MHz) &  -238.7 & -239.6 & -238.2 & -240.0 & -233.6 & -273.8 & -244.8 & -241.4 & -241.1 \\

$\omega_{\mathrm{r}}/2\pi$ (GHz) & 7.221 & 7.170 & 7.121 & 7.304 & 6.942 & 7.324 & 7.208 & 7.28 & 7.155    \\

$f_{\mathrm{gg}}$ & 0.98 & 0.99 & 0.99 & 0.99 & 0.99 & 0.99 & 0.99 & 0.93 & 0.99 \\

$f_{\mathrm{ee}}$ & 0.93 & 0.87 & 0.94 & 0.94 & 0.95 & 0.94 & 0.94 & 0.89 & 0.94 \\

Readout assignment fidelity & 0.95 & 0.93 & 0.96 & 0.97 & 0.97 & 0.96 & 0.96 & 0.91 & 0.96  \\

$T_1^{\rm avg}$ ($\SI{}{\micro s}$)&  11.2 & 10.9 & 11.1 & 11.6 & 12.1 & 10.1 & 11.9 & 11.3 & 10.7  \\

\hline
\hline
\end{tabular}
}
\end{table}

Detailed sample parameters for individually isolated qubits are summarized in Tab.~\ref{tab:sample_parameters}. The samples are fabricated on a silicon substrate by dry etching an MBE grown, \SI{250}{nm} thick aluminum film in an optical lithography process, forming all larger circuit elements such as the qubit capacitor pads, resonators and the signal lines for qubit readout and control. The qubit SQUID loops are fabricated with an electron beam lithography process and a double-angle shadow evaporation technique~\cite{Dolan1977} to form the Josephson junctions. We use a $\SI{5}{\micro m}$ wide wire width for the SQUIDs, which was found to minimize flux noise from local magnetic spin defects on the SQUID surface and interfaces~\cite{Braumueller2020}.

Our experimental setup is shown in Fig.~\ref{fig:setup}, with the sample mounted to the base stage of a dilution refrigerator at a temperature of approximately $\SI{20}{mK}$. Our readout setup follows the established heterodyne mixing scheme. The reflected microwave signal is amplified with a travelling wave parametric amplifier (TWPA)~\cite{Macklin2015}, which is crucial for our experiment due to its large bandwidth and high saturation level. We use a heterodyne mixing scheme~\cite{Braumuller2021,Krantz2019} for qubit XY-control,  enabling the application of simultaneous qubit control pulses by frequency-selecting mutually detuned qubits. DC and fast flux control pulses are combined in an RF-choke before they reach the on-chip flux bias lines.

\section{Pulse sequences}

\begin{figure*}
\includegraphics{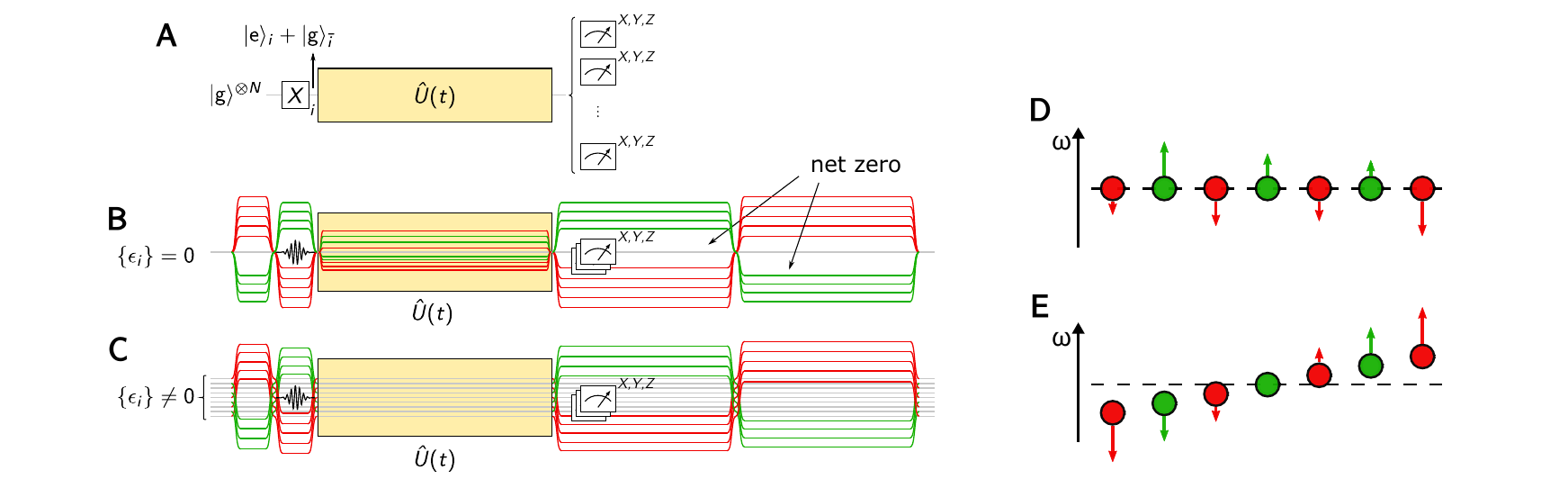}
\caption{
\textbf{Pulse sequences used in experiments}
\textbf{(A)} General circuit diagram of the pulse sequences used in our experiments. With all qubits initially in the ground state $\Ket{\mathrm{g}}$, we inject one particle into the lattice by exciting the qubit at site $i$ with a $X$ gate. After the unitary time evolution $\hat U (t)$ for time $t$, we perform simultaneous readout of all $N=9$ qubits.
\textbf{(B)} Pulse sequence with fast flux pulses, which tune the qubit frequencies, shown in red and green. Here, we apply dc-fluxes such that all qubits are on resonance, $\epsilon_i=0$. Prior to the unitary evolution, we mutually detune the qubits from each other via net-zero flux pulses and apply the $X$ gate on qubit $i$. During free, unitary evolution $\hat U (t)$ the lattice is either degenerate ($\epsilon_i=0)$ or we can apply some frequency detuning via another set of fast flux pulses. Finally, we again detune qubits from each other and perform simultaneous readout.
\textbf{(C)} In contrast to the sequence in (B), we apply the desired frequency detunings $\epsilon_i\neq 0$ via the dc-flux signals. We again detune the qubits from each other for state preparation and readout but do not require additional flux pulses during the unitary evolution. \textbf{(D)} Checkerboard stagger frequency detuning scheme. We use this scheme for state preparation and readout when emulating uniform and disordered lattices. \textbf{(E)} Ladder frequency detuning scheme. We use this scheme for state preparation and readout when emulating lattices with a potential gradient.
}
\label{fig:pulse_sequences}
\end{figure*}

The general pulse sequence for our experiments is depicted in Fig.~\ref{fig:pulse_sequences}A. With all $N$ qubits initially in the ground state, $\Ket{\mathrm{g}}^{\otimes N}$, we inject one particle into the lattice by exciting the qubit at site $i$ with a $\pi$-pulse around the $X$-axis of the Bloch sphere (Pauli-$X$ gate). After the unitary time evolution $\hat U (t)$ we read out the states of all $N=9$ qubits simultaneously. We map the measured qubit states to the expectation values of $\hat\sigma^z_i$ by using individual $\beta$-matrices for each qubit which we have separately calibrated via Rabi measurements~\cite{Kjaergaard2020}. In order to extract information about the qubit states along all three axes of the Bloch sphere, we prepend the readout pulse with appropriate tomography rotation pulses. 

In order to extract the concurrence data in Fig.~2 of the main text, we reconstruct the two-qubit density matrices of the respective qubit pairs. We reconstruct the two-qubit density matrices via two-qubit state tomography, using a maximum-likelihood estimation~\cite{Banaszek2000} based on nine successive tomography measurements in the respective two-qubit subspace.

Schematic pulse sequences used in our experiments are shown in Fig.~\ref{fig:pulse_sequences}B, C. Fast flux pulses which tune the qubit frequencies $\epsilon _i$ are shown in green (even numbered qubits) and red (odd numbered qubits). We mutually detune the qubits prior to the unitary evolution to enable state preparation, and after the unitary evolution for qubit readout. Both tuning pulses are performed in net-zero fashion~\cite{Rol2019} in order to cancel long-timescale transient distortions.

For our experiments that require a degenerate lattice ($\epsilon _k=0$), we apply dc-fluxes such that all qubits are on resonance, and during the unitary evolution we apply no additional fast flux pulses. In order to apply a controlled frequency disorder ($\epsilon _k\neq 0$) during the unitary evolution we use two variants: in Fig.~\ref{fig:pulse_sequences}B, the dc-fluxes initially tune all qubits on resonance. We then apply additional fast flux pulses during unitary evolution to realize the desired frequency detunings. In the alternative variant (Fig.~\ref{fig:pulse_sequences}C), we apply the desired frequency detunings via the dc-flux signals, yielding the desired frequency detunings during $\hat U (t)$ without the need for any further fast flux pulsing. The variant in Fig.~\ref{fig:pulse_sequences}B has the advantage that the qubit frequencies are independent of the experimental parameters during excitation and readout, such that we do not have to recalibrate these steps when sweeping the disorder $\delta/J$ or Wannier-Stark potential $F/J$. Data in Fig.~3, 4 for the 2d lattice were therefore measured with the variant in Fig.~\ref{fig:pulse_sequences}B, while the simpler 1d configurations were measured with the variant shown in Fig.~\ref{fig:pulse_sequences}C.

In the 1d chain, we apply the Wannier-Stark potential with potential slope $F$ according to $\epsilon_k=kF$, where $\epsilon _k$ is the qubit frequency detuning of site $i$ relative to the frame rotating at $\omega_{\mathrm{ref}}$ and $k=0$ is the site in the center of the chain. In the 2d lattice, we apply the Wannier-Stark potential along the propagation direction (diagonal) of the lattice, such that qubits at the same Manhattan distance from the corner qubit receive an identical detuning frequency. For the measurements with anisotropic Wannier-Stark potential, we apply two different potential slopes $F_x \neq F_y$ along two perpendicular axes $x$ and $y$ of the lattice.

In order to prevent unwanted interaction between qubits during state preparation and readout we ensure that neighboring qubits are detuned by at least $\SI{180}{MHz}$ ($\approx 22.5 J$). When emulating uniform and disordered lattices, we realize this requirement using a checkerboard stagger (Fig.~\ref{fig:pulse_sequences}D) where we detune the odd-numbered qubits down in frequency and the even-numbered qubits up in frequency using fast-flux pulses. In contrast, when emulating potential gradients, we use a ladder stagger (Fig.~\ref{fig:pulse_sequences}E) where we detune the qubits according to the lattice gradient in a manner that prevents the frequency neighboring qubits from becoming on-resonance during the process. We have found that using a ladder stagger is essential in reducing the errors in emulating Bloch oscillations and Wannier-Stark localization resulting from our pulse sequence.

\section{Simulations}

To verify our experimental results, we simulate the dynamics of the system. We construct the emulated Hamiltonian using the measured couplings between neighboring qubits and the desired detunings in the rotating frame. We include the non-unitary terms arising from qubit relaxation and dephasing using the Lindblad master equation
\begin{align}
\begin{split}
    \dot{\rho}(t)=&-\frac{i}{\hbar}[\hat{H},\rho(t)] \\
    &+\frac{\gamma_r}{2}\sum_i \big( 2\sigma^-_i \rho(t) \sigma^+_i - \rho(t) \sigma^+_i \sigma^-_i - \sigma^+_i \sigma^-_i \rho(t) \big) \\
    &+\gamma_\phi\sum_i \big( \sigma^z_i \rho(t) \sigma^z_i -  \rho(t) \big),
\end{split}
\end{align}
where $\sigma^-$ ($\sigma^+$) is the qubit annihilation (creation) operator, and $\sigma^z$ is the Pauli-$Z$ operator. The average qubit relaxation rate $\gamma_r=1/T_1$ is measured, and the average dephasing rate $\gamma_\phi=1/T_\phi$ is set to a value that best describes the experimental data. We note that the $\gamma_\phi$ used in our simulations is smaller than the measured value at the operation frequency by a factor of $\approx 3$. The reduction in the dephasing rate is due to the reduced flux dispersion in the coupled qubits spectrum, which makes the system less sensitive to magnetic flux noise.

\section{2d quantum random walk in a $3 \times 3$ lattice}

\begin{figure}[h!]
\subfloat{\label{fig:3x3_lattice_full}}
\subfloat{\label{fig:3x3_lattice_reduced}}
\includegraphics{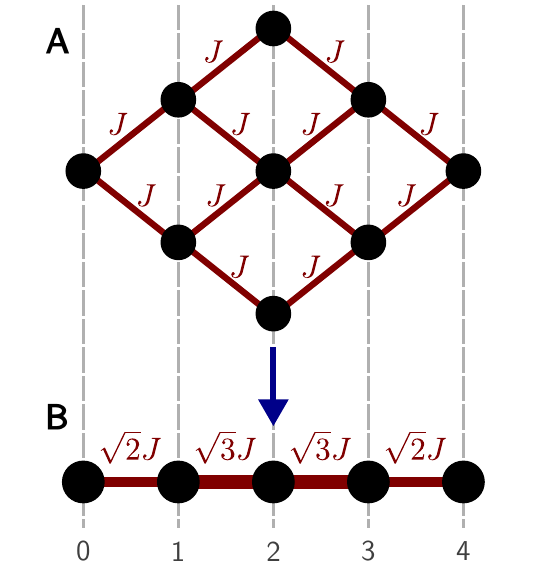}
\caption{
\textbf{$3 \times 3$ lattice} \textbf{(A)} The lattice can be represented by graph where the sites are represented by nodes and the couplings are represented by edges. \textbf{(B)} The quantum random walk on a $3 \times 3$ graph starting from a corner can be reduced to a walk on a line. The nodes of the reduced graph represent the qubits at the same distance from the start point, and the edges reflect the effective couplings.}
\label{fig:3x3_lattice}
\end{figure}

A 2d tight-binding lattice can be treated as a graph with the lattice sites as nodes, and the nearest-neighbor couplings as edges. Assuming a single particle propagation in the lattice, we represent the quantum state of the system with the particle on node $a$ as $\ket{a}$. Hence, the tight-binding Hamiltonian $H$ described in Eq.~1 contains the following matrix elements: 
\begin{equation}
  \bra{a}H\ket{b} =
    \begin{cases}
      -J & \text{if $a$ and $b$ are neighbors}\\
      0 & \text{otherwise}
    \end{cases}       
\end{equation}
Using this notion, we can show that in a 1d graph, 
\begin{equation}
    H\ket{j}=-J(\ket{j-1}+\ket{j+1})
\end{equation}

We now analyze the QRW on a 2d graph of our $3 \times 3$ lattice starting in the state corresponding to one of the corners (Fig.~\ref{fig:3x3_lattice_full}). This subspace is spanned by states $\ket{\rm{col} \: j}$, defined as the uniform superposition over all vertices at distance $j$ away from the root~\cite{Childs2001}:
\begin{equation}
    \ket{\rm{col} \: j} = \frac{1}{\sqrt{N_j}} \sum_{a \in \text{distance j}}\ket{a},
\end{equation}
where $N_j$ is the total number of nodes at distance $j$. In this basis,
\begin{align}
\begin{split}
    \bra{\rm{col} \: 0} H \ket{\rm{col} \: 1} &= -\sqrt{2}J\\
    \bra{\rm{col} \: 1} H \ket{\rm{col} \: 2} &= -\sqrt{3}J\\
    \bra{\rm{col} \: 2} H \ket{\rm{col} \: 3} &= -\sqrt{3}J\\
    \bra{\rm{col} \: 3} H\ket{\rm{col} \: 4} &= -\sqrt{2}J
\end{split}
\end{align}

Based on this analysis, we can represent the 2d QRW starting from the corner of a $3 \times 3$ lattice as a QRW on a line with $5$ nodes, where each node is a superposition of the sites at the same distance from the graph root, and the coupling strengths between the nodes vary (Fig.~\ref{fig:3x3_lattice_reduced}). The average coupling strength on this chain is $J^{\text{avg}}_{\text{eff}}=(\sqrt{2}+\sqrt{3})J/2$, resulting in an average particle propagation velocity $v_g=\sqrt{2}J^{\text{avg}}_{\text{eff}}=(1+\sqrt{3/2})J$~\cite{Konno2005,Scardicchio2017}.

\section{Entanglement propagation}

\begin{figure}[h!]
\subfloat{\label{fig:concurrence_avg}}
\subfloat{\label{fig:concurrence_7_all}}
\includegraphics{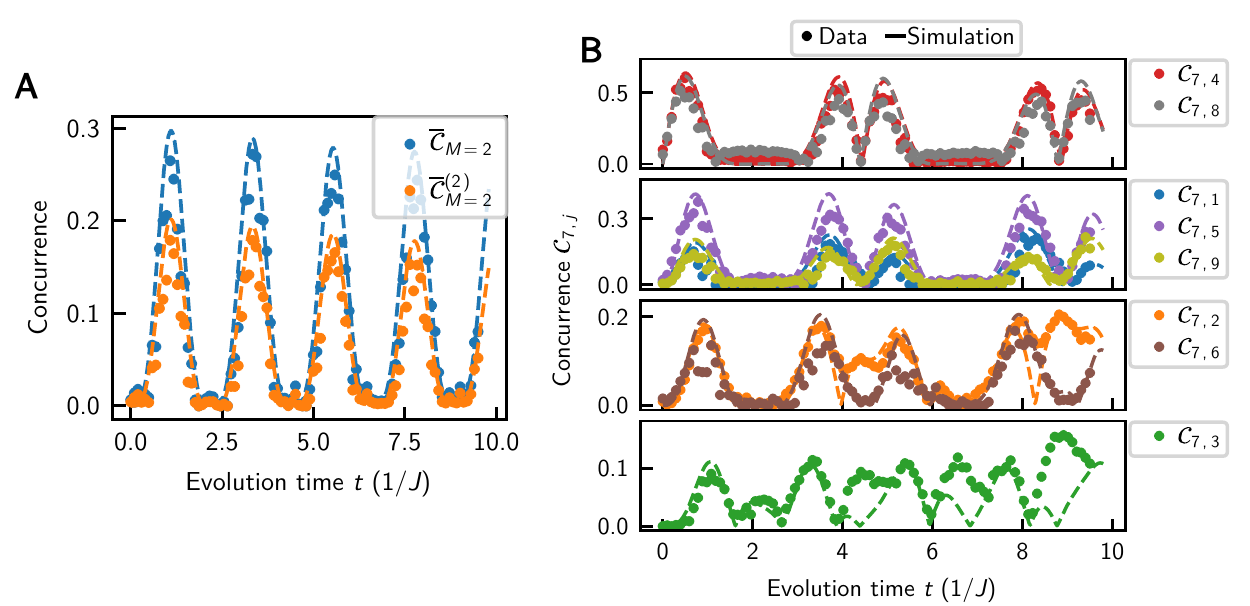}
\caption{
\textbf{Concurrence} \textbf{(A)} Distributed entanglement formed among the three qubits at distance $M=2$ from the propagation starting point. $\mathcal{C}^{\text{min}}_{i,(j,k)}$ represents the minimum concurrence between qubit $i$ and the sub-system consisting of qubits $j$ and $k$. \textbf{(B)} Comparing the sub-system average concurrence $\bar{\mathcal{C}}$ defined in Eq.~\ref{eq:avg_conc} with the average pairwise concurrence $\bar{\mathcal{C}}^{(2)}$ among the sites at distance $M=2$. \textbf{(C)} Pairwise concurrence between qubit $7$ and each other qubit in the lattice.}
\label{fig:concurrence}
\end{figure}

In this work we use concurrence as the measure for the entanglement between two sub-systems. The entanglement of formation $E(\rho)$ of a bipartite mixed state described by $\rho$ can be related to concurrence~\cite{Wootters1998}:
\begin{equation}
    E(\rho)=h\left(\frac{1+\sqrt{1-C(\rho)^2}}{2}\right)
\end{equation}
where $h(x)=-x\:\mathrm{log}(x)-(1-x)\:\mathrm{log}(1-x)$ is the binary entropy function.

Concurrence can be expanded to explore the distributed entanglement between three, or more, qubits~\cite{Coffman2000}. For three qubits $A$, $B$, and $C$, the entanglement between $A$ and a sub-system consisting of the pair $BC$ can be quantified using $\mathcal{C}_{A,(B,C)}$. In \cite{Coffman2000} it is shown that 
\begin{equation}
    \mathcal{C}_{A,B}^2 + \mathcal{C}_{A,C}^2 \leq \mathcal{C}_{A,(B,C)}^2.
\label{eq:concurrence_inequality}
\end{equation}
For an arbitrary single-particle state $\ket{\phi}=\alpha \ket{100}+\beta \ket{010}+\gamma \ket{001}$ the inequality in Eq.~\ref{eq:concurrence_inequality} turns into an equality~\cite{Coffman2000}: For $\ket{\phi}$, $\mathcal{C}_{A,B}=2|\alpha \beta|$, $\mathcal{C}_{A,B}=2|\alpha \gamma|$, and $\mathcal{C}_{A,(B,C)}=2|\alpha|\sqrt{|\beta|^2+|\gamma|^2}$. Hence, $\mathcal{C}_{A,B}^2 + \mathcal{C}_{A,C}^2 = \mathcal{C}_{A,(B,C)}^2$. For a generalized $N$ qubit state $\ket{\phi}= \sum_\pi \alpha_\pi \ket{\pi (0...01)}$, it can be shown that~\cite{Coffman2000}:
\begin{equation}
    \sum^N_i \mathcal{C}^2_{1,i}=\mathcal{C}^2_{1,(2,3,...,N)},
\label{eq:concurrence_W_equality}
\end{equation}
where we label the qubits by numbers instead of letters.

In our tight-binding lattice, a sub-system consisting of three sites, when considered independent of the rest of the system, is in a mixed state. In this scenario, Eq.~\ref{eq:concurrence_inequality} describes the bound for the minimum concurrence $(\mathcal{C}^2)^{\rm min}_{A,(B,C)}$ over all decompositions of the three-qubit density matrix. In Fig.~2B of the main text we show the time evolution of the lower-bound of the distributed concurrence between the qubits at distance $M=2$, calculated using the pairwise concurrence values reported in Fig.~2A of the main text. For this sub-system we define the average concurrence $\bar{\mathcal{C}}$ as
\begin{equation}
    \bar{\mathcal{C}}^2=\frac{1}{3} \big[(\mathcal{C}^2)^{\rm min}_{1,(5,9)} + (\mathcal{C}^2)^{\rm min}_{5,(1,9)} + (\mathcal{C}^2)^{\rm min}_{9,(1,5)} \big].
\label{eq:avg_conc}
\end{equation}
For lattice sub-systems with only two qubits labeled by $i$ and $j$, the average concurrence is simply the pairwise concurrence value: $\bar{\mathcal{C}}=\mathcal{C}_{i,j}$. In Fig.~\ref{fig:concurrence_avg} we compare $\bar{\mathcal{C}}$ to the average pairwise concurrence $\bar{\mathcal{C}}^{(2)}$ for the sites at $M=2$. While both quantities show a similar qualitative evolution, $\bar{\mathcal{C}}$ can better quantify the entanglement in this sub-system by accounting for the distributed entanglement in a three-qubit system.

In Fig.~2D we show the concurrence between the propagation source (site $7$) of the QRW and the rest of the lattice. This quantity is extracted using the relation described in Eq.~\ref{eq:concurrence_W_equality} and the measured pairwise concurrence values between site $7$ and the other sites in the lattice shown in Fig.~\ref{fig:concurrence_7_all}.

\section{Quantum random walks}

\begin{figure}[h!]
\subfloat{\label{fig:1d_QRW_middle}}
\subfloat{\label{fig:1d_x2_middle}}
\subfloat{\label{fig:QRW_fidelity}}
\includegraphics{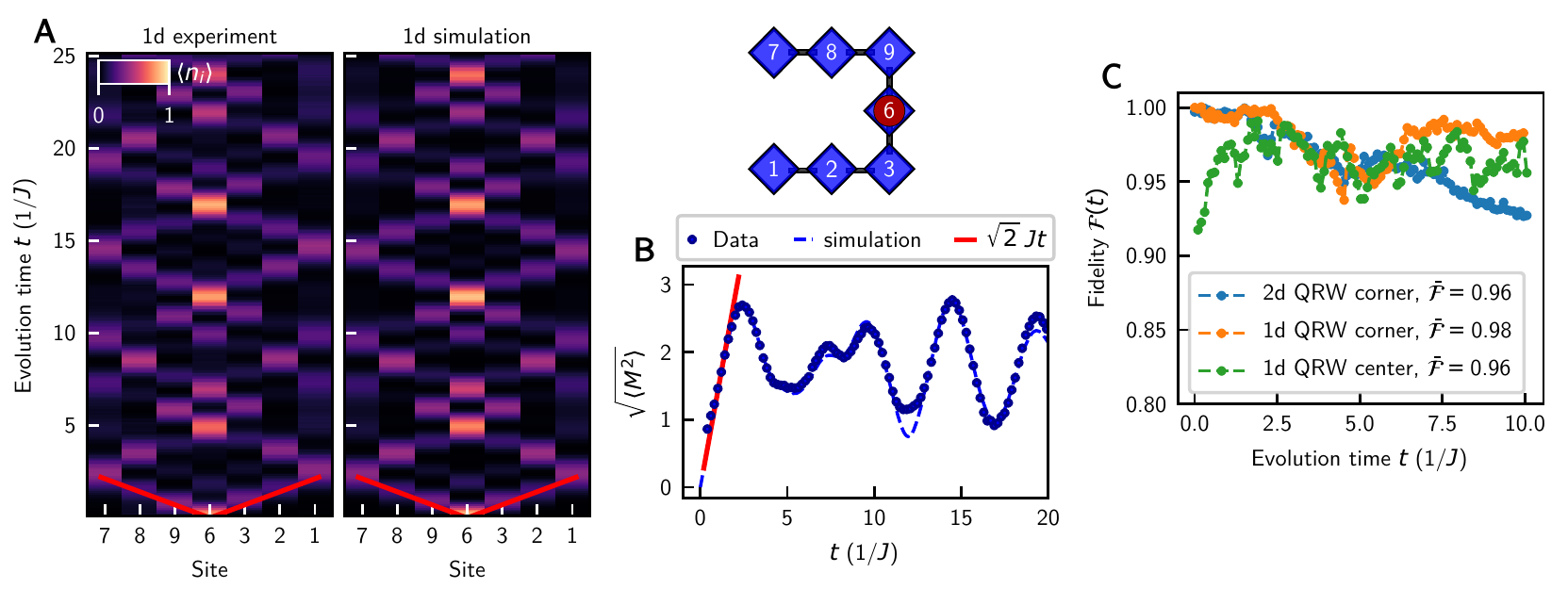}
\caption{
\textbf{Quantum random walks} \textbf{(A)} 1d quantum random walk with a particle initialized at the center of the chain (site 6). \textbf{(B)} The evolution of the second moment of position in time. We observe ballistic propagation with propagation speed $v_g=\sqrt{2} J$. \textbf{(C)} Fidelity $\mathcal{F}(t)$ of the experimental compared to ideal numerical simulations. During the first $\SI{200}{ns}$ of evolution we observe an average fidelity $\bar{\mathcal{F}}>96\%$}
\label{fig:QRW}
\end{figure}

In our experiments we realize quantum random walks a 1d lattice consisting of 7 qubits, and a 2d lattice consisting of 9 qubits. We initialize the system on resonance using our dc-flux control, and then detune the qubits using fast-flux pulses followed by exciting one of the qubits with a $X_\pi$ pulse. Afterward all qubits are brought on resonance (their idle point), and the system is allowed to evolve for time $t$. Afterwards, we detune the qubits once again using fast-flux pulse to halt the dynamics during the simultaneous readout of the qubits. In Fig.~1D and Fig.~1E we show a 2d random walk, and a 1d random walk with the particle injected at the edge of the 7-qubit chain. In Fig.~\ref{fig:1d_QRW_middle} we demonstrate a 1d random walk with the particle initialized at the center of the chain. By evaluating the particle's second moment of position $\langle M^2 \rangle$ (Fig.~\ref{fig:1d_x2_middle}) we are able to observe that in this scenario the particle propagates with the group velocity $v_g=\sqrt{2} J$ prior to reflection from the lattice boundaries, which is in agreement with theory~\cite{Hoyer2010,Scardicchio2017}. 

In order to evaluate the quality of our QRW, we compare the result of our experiments with ideal numerical simulations using the fidelity
\begin{equation}
    \mathcal{F}(t)=\sum_i \sqrt{p_i(t)q_i(t)}
\end{equation}
where $p_i(t)$ and $q_i(t)$ are the measured and theoretical probability distributions for site $i$ at time $t$ respectively. The high fidelity values reported in Fig.~\ref{fig:QRW_fidelity} exhibit a high degree of control in realizing coherent 1d and 2d random walks.

\section{Relation between participation ratio and localization length}

The spatially localized wavefunction with localization length $\xi$ on a lattice takes the form
\begin{align}
\begin{split}
    \psi^{\rm 1d}(x_i)&=e^{-|x_i|/\xi}, \\
    \psi^{\rm 2d}(x_i,y_i)&=e^{-|x_i+y_i|/\xi}.
\end{split}
\end{align}
where $x_i$ and $y_i$ are the horizontal and vertical position of site $i$ on the lattice. Using this wavefunction form we can derive an analytical form for the participation ration as a function of $\xi$ for an infinite 1d chain with the propagation source at the center $PR^{\rm 1d}_{\infty}$, and the edge $PR^{\rm 1d}_{\infty, \rm edge}$:
\begin{align}
\begin{split}
    PR^{\rm 1d}_\infty (\xi)&= \big(\sum_{i=-\infty}^\infty e^{-2|x_i|/\xi} \big)^2/ \big(\sum_{i=-\infty}^\infty e^{-4|x_i|/\xi} \big) = 2 \: \rm{coth} \big(1/\xi \big) -  \rm{tanh} \big(2/\xi \big)\\
    PR^{\rm 1d}_{\infty, \rm edge} (\xi) &= \big(\sum_{i=0}^\infty e^{-2|x_i|/\xi} \big)^2/ \big(\sum_{i=0}^\infty e^{-4|x_i|/\xi} \big) = \rm{coth} \big(1/\xi \big)
\end{split}
\end{align}

In a finite lattice the relations above are altered, although they can still be extracted analytically. For a lattice with $N$ sites, where the propagation source at one end of the chain 
\begin{equation}
    PR^{\rm 1d}_{N, \rm edge} (\xi) = \big(\sum_{i=0}^{N-1} e^{-2|x_i|/\xi} \big)^2/ \big(\sum_{i=0}^{N-1} e^{-4|x_i|/\xi} \big) = \rm{coth(1/\xi)} \: \rm{tanh} (N/\xi).
\end{equation}
In the full localization ($\xi \rightarrow 0$) $PR_N^{\rm edge} \rightarrow 1$, and for a fully extended state ($\xi \rightarrow \infty$) $PR_N^{\rm edge} \rightarrow N$ which are compatible with expected values of the participation ratio.

The same analysis can also be extended to a 2d lattice. For a propagation source at the corner of an infinite lattice 
\begin{equation}
    PR^{\rm 2d}_{\infty, \rm corner} (\xi) = \big(\sum_{i,j=0}^\infty e^{-2|x_i+y_j|/\xi} \big)^2/ \big(\sum_{i,j=0}^\infty e^{-4|x_i+y_j|/\xi} \big) = \rm{coth}^2 \big(1/\xi \big)
\end{equation}
For a finite $n \times n$ lattice, we find $PR^{\rm 2d}_{n \times n, \rm corner}$ by truncating the sum above:
\begin{equation}
    PR^{\rm 2d}_{n \times n, \rm corner} (\xi) = \big(\sum_{i,j=0}^{n-1} e^{-2|x_i+y_j|/\xi} \big)^2/ \big(\sum_{i,j=0}^{n-1} e^{-4|x_i+y_j|/\xi} \big) =  \rm{coth}^2(1/\xi) \: \rm{tanh}^2 (n/\xi)
\end{equation}

\pagebreak

\section{Extended data}

\begin{figure}[h!]
\includegraphics{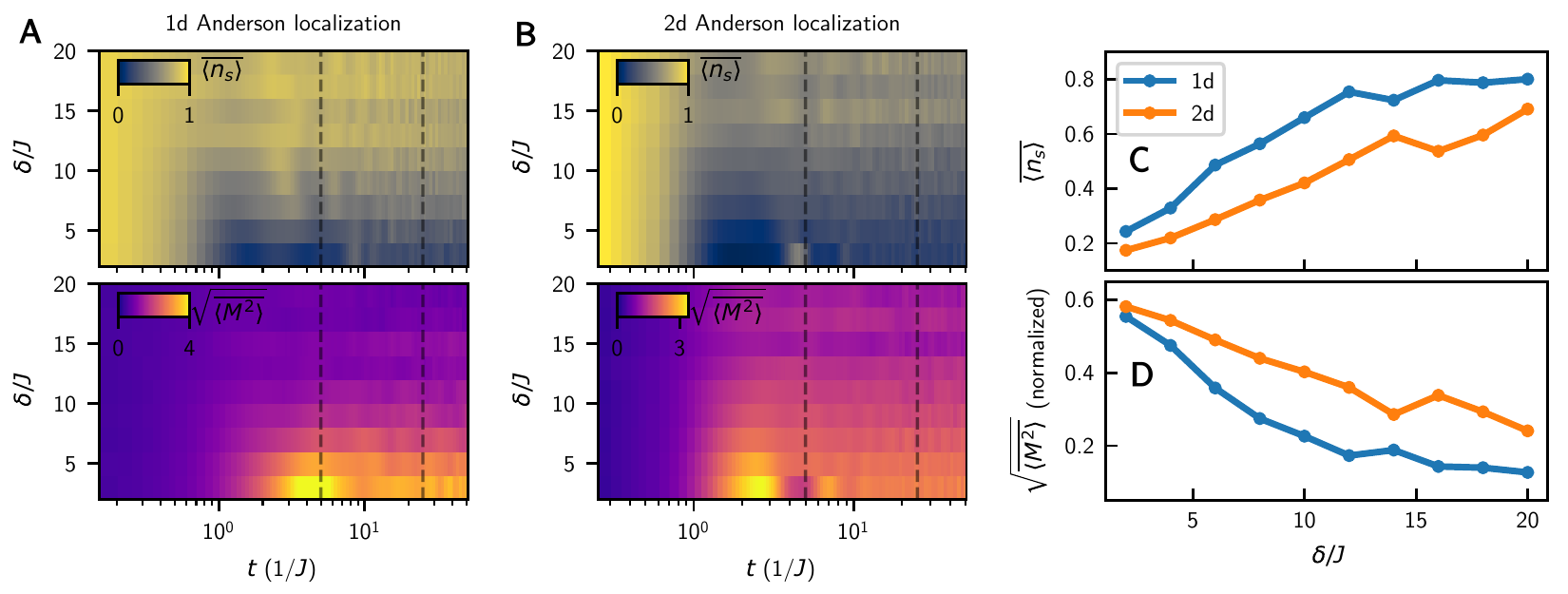}
\caption{
\textbf{Anderson Localization} Time evolution of the average population on the particle initialization site ($\overline{\langle n_s \rangle}$) and the particle spread  ($\sqrt{\overline{\langle M^2 \rangle}}$) averaged over $60$ disorder realizations for each disorder strength $\delta$  in a \textbf{(A)} 7-site chain and \textbf{(B)} a $3 \times 3$ lattice. Time-averaged \textbf{(C)} $\overline{\langle n_s \rangle}$ and \textbf{(D)} $\sqrt{\overline{\langle M^2 \rangle}}$ (normalized to the maximum value allowed by the lattice size) in the time-span $5/J < t < 25/J$ (marked by dashed lines) are reported for the different values of $\delta$.}
\label{fig:Anderson_extended}
\end{figure}

\begin{figure}[h!]
\includegraphics{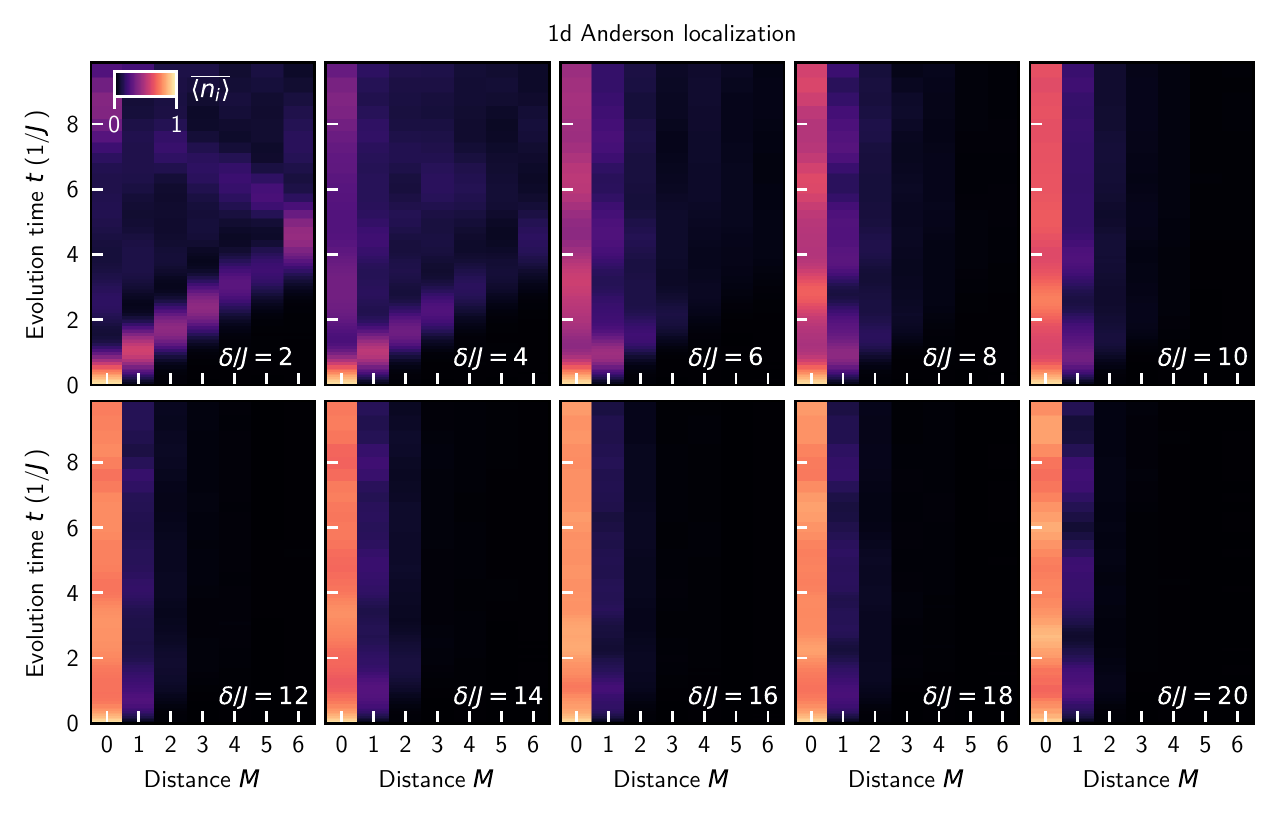}
\caption{
\textbf{1d Anderson Localization} The time evolution of the probability density at each site averaged over $60$ random lattice realizations for each disorder strength $\delta$.}
\label{fig:Anderson_1d_profile}
\end{figure}

\begin{figure}[h!]
\includegraphics{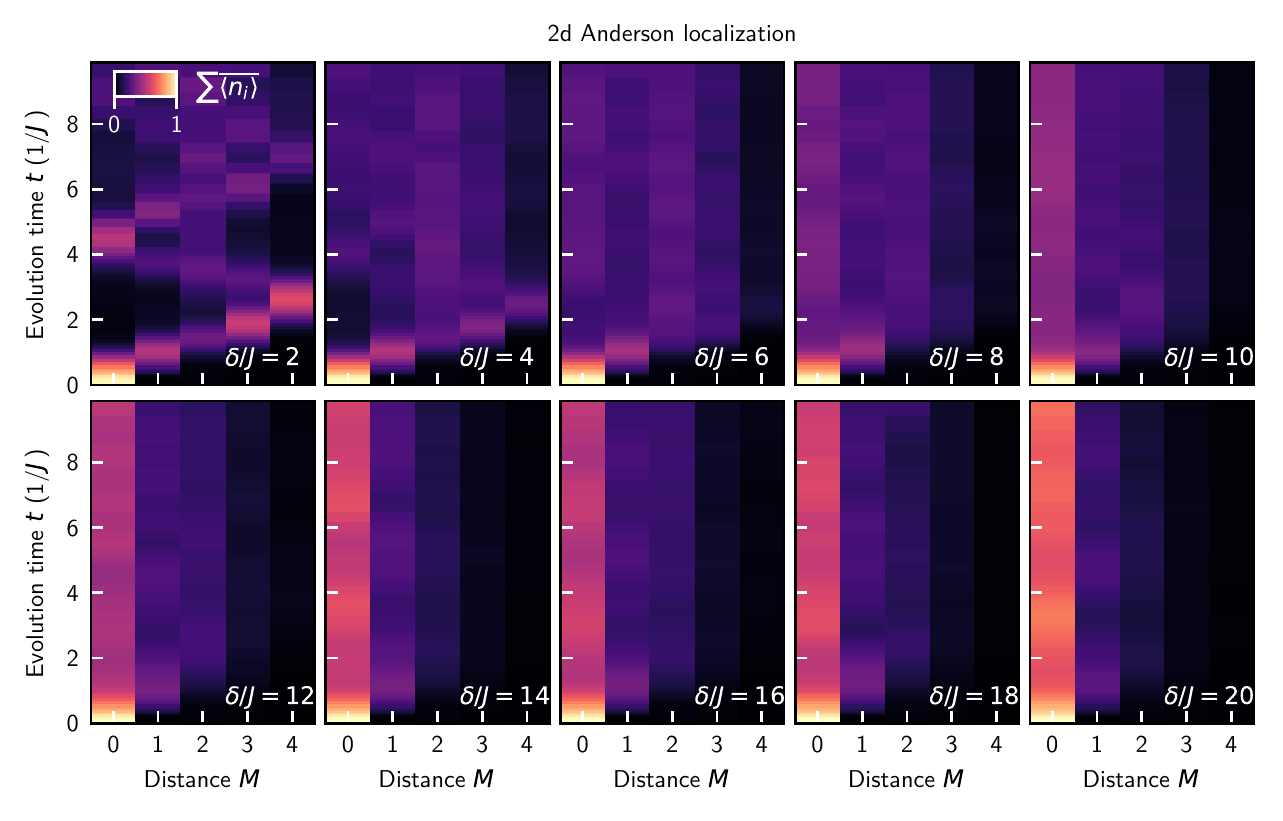}
\caption{
\textbf{2d Anderson Localization}  The time evolution of the probability density at each distance from the propagation sources averaged over $60$ random lattice realizations for each disorder strength $\delta$.}
\label{fig:Anderson_2d_profile}
\end{figure}

\begin{figure}[h!]
\includegraphics{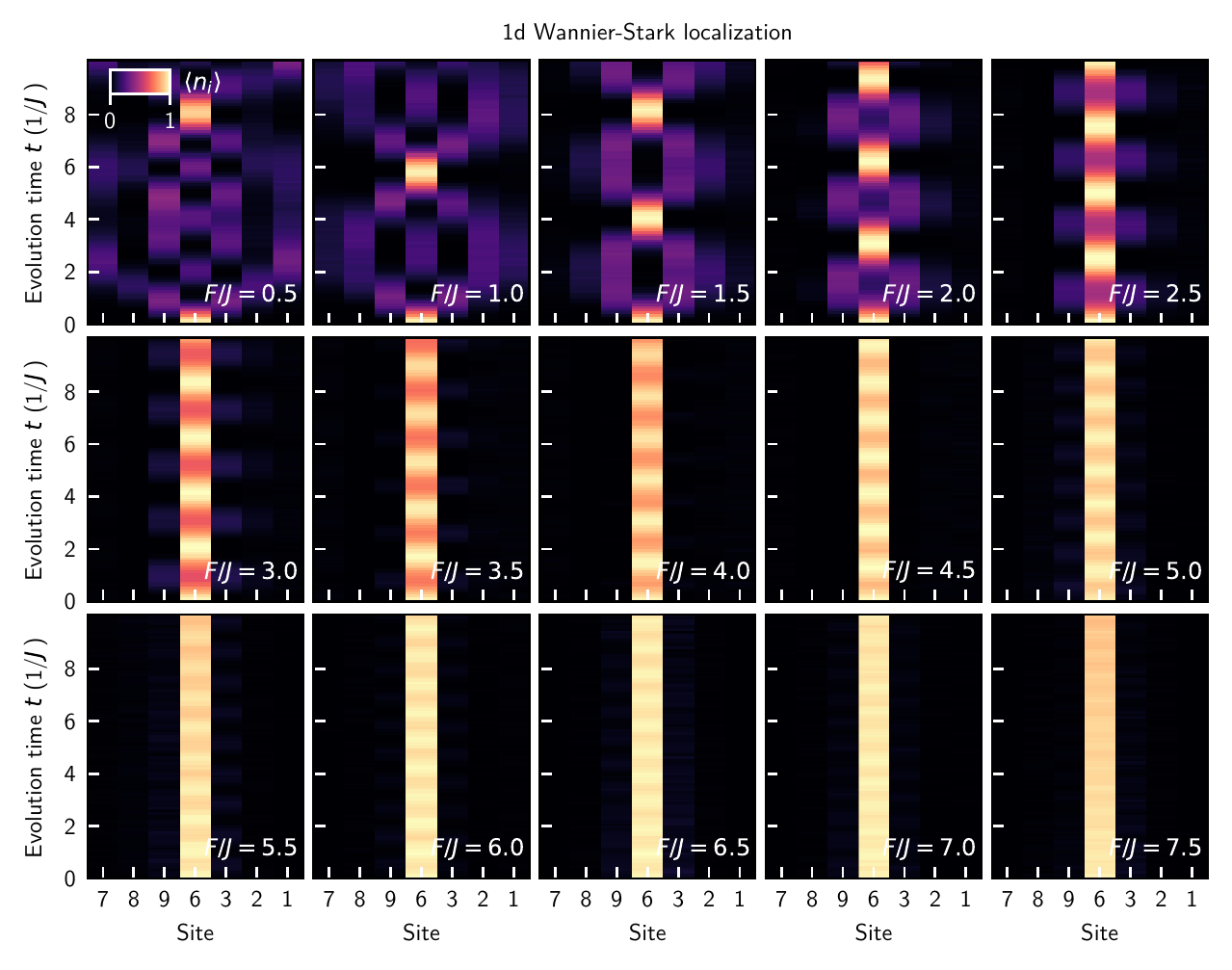}
\caption{
\textbf{1d Wannier-Stark localization} The time evolution of the probability density at each site for different potential gradient values $F$.}
\label{fig:Stark_1d_profile}
\end{figure}

\begin{figure}[h!]
\includegraphics{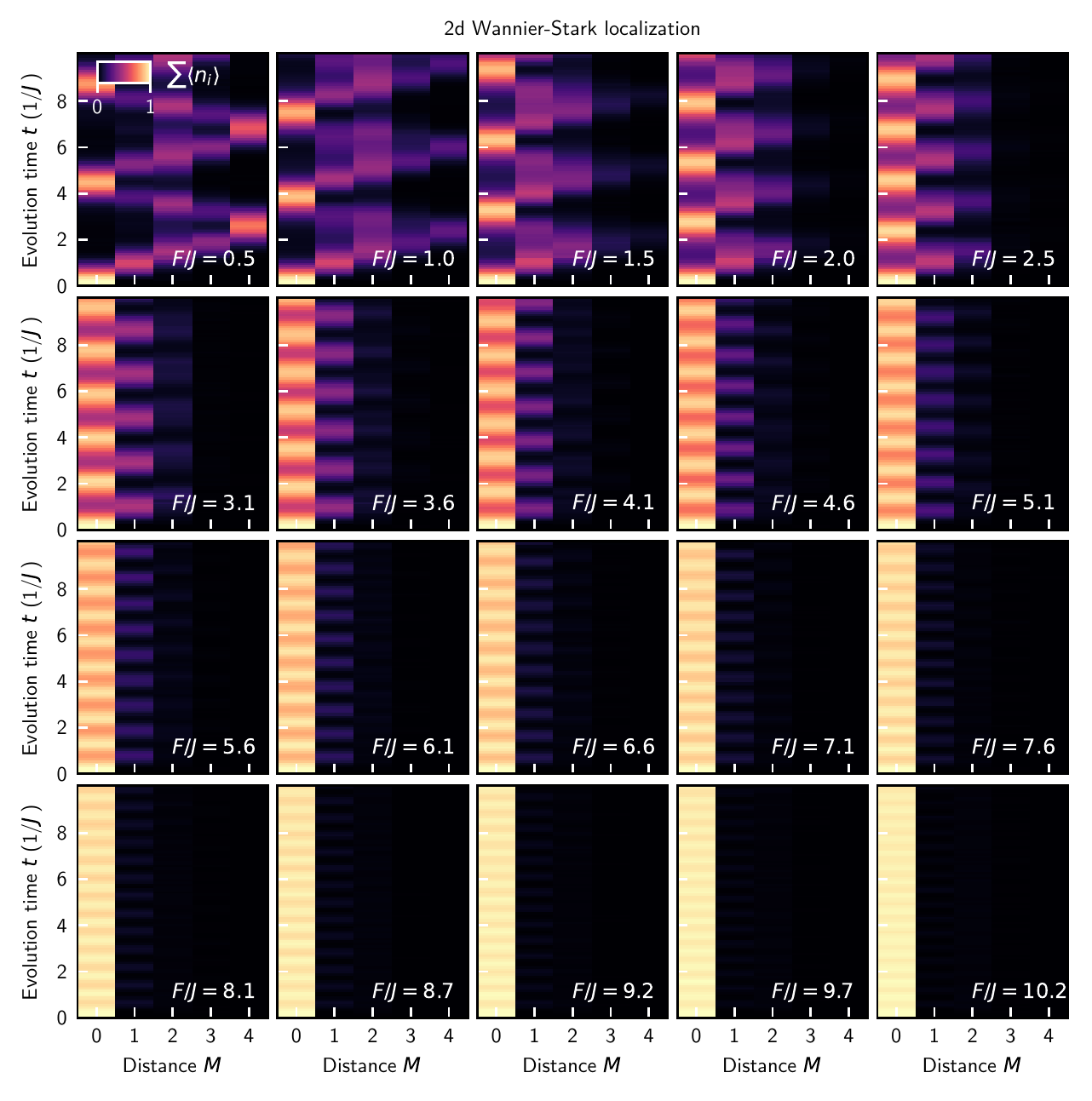}
\caption{
\textbf{2d Wannier-Stark localization profile} The time evolution of the probability density at each distance from the propagation source for different isotropic potential gradients with strength $F$ along each axis.}
\label{fig:Stark_2d_profile}
\end{figure}

\clearpage

\bibliography{supplement}